\documentclass[a4paper,12pt]{article}
\usepackage{authblk} 

\usepackage[english]{babel}
\usepackage[utf8]{inputenc}
\usepackage{amsmath,amsfonts,amssymb}
\usepackage{float}
\usepackage[margin=1in]{geometry}
\usepackage{graphicx}
\usepackage[hidelinks]{hyperref}
\usepackage[super,comma]{natbib}

\DeclareMathOperator{\logit}{logit}
\newcommand{\indep}{\raisebox{0.05em}{\rotatebox[origin=c]{90}{$\models$}}}
\newcommand*\given{\,|\,}
\newcommand*\diff{\mathop{}\!\mathrm{d}}

\title{Multiple imputation of missing covariates when using the Fine--Gray model}

\author[1]{Edouard F. Bonneville} 
\author[2]{Jan Beyersmann} 
\author[3]{Ruth H. Keogh} 
\author[3]{Jonathan W. Bartlett} 
\author[4]{Tim P. Morris} 
\author[5]{Nicola Polverelli} 
\author[1,6,*]{Liesbeth C. de Wreede} 
\author[1,7,*]{Hein Putter} 

\affil[1]{\footnotesize Department of Biomedical Data Sciences, Leiden University Medical Center, the Netherlands}
\affil[2]{\footnotesize Institute of Statistics, Ulm University, Germany}
\affil[3]{\footnotesize Department of Medical Statistics, London School of Hygiene and Tropical Medicine, United Kingdom}
\affil[4]{\footnotesize MRC Clinical Trials Unit at UCL, United Kingdom}
\affil[5]{\footnotesize Unit of Bone Marrow Transplantation, Division of Hematology, Fondazione IRCCS Policlinico San Matteo di Pavia, Italy}
\affil[6]{\footnotesize DKMS Clinical Trials Unit, Germany}
\affil[7]{\footnotesize Mathematical Institute, Leiden University, the Netherlands}
\affil[*]{Shared senior authorship}

\date{}

\begin{document}

\maketitle

\begin{abstract}
The Fine--Gray model for the subdistribution hazard is commonly used for estimating associations between covariates and competing risks outcomes. When there are missing values in the covariates included in a given model, researchers may wish to multiply impute them. Assuming interest lies in estimating the risk of only one of the competing events, this paper develops a substantive-model-compatible multiple imputation approach that exploits the parallels between the Fine--Gray model and the standard (single-event) Cox model. In the presence of right-censoring, this involves first imputing the potential censoring times for those failing from competing events, and thereafter imputing the missing covariates by leveraging methodology previously developed for the Cox model in the setting without competing risks. In a simulation study, we compared the proposed approach to alternative methods, such as imputing compatibly with cause-specific Cox models. The proposed method performed well (in terms of estimation of both subdistribution log hazard ratios and cumulative incidences) when data were generated assuming proportional subdistribution hazards, and performed satisfactorily when this assumption was not satisfied. The gain in efficiency compared to a complete-case analysis was demonstrated in both the simulation study and in an applied data example on competing outcomes following an allogeneic stem cell transplantation. For individual-specific cumulative incidence estimation, assuming proportionality on the correct scale at the analysis phase appears to be more important than correctly specifying the imputation procedure used to impute the missing covariates.
\end{abstract}

{\bf Keywords:} Fine--Gray model; subdistribution hazard; competing risks; multiple imputation; missing covariates; cumulative incidence function

\section{Introduction}

The presence of missing covariate data continues to be pervasive across biomedical research. Among the many existing approaches for dealing with missing covariate data, multiple imputation (MI) methods in particular have become increasingly popular in practice \citep{carpenterMissingDataStatistical2021a}. Compared to a complete-case analysis (CCA), MI can provide inferences that are both less biased and more efficient, under certain missingness mechanisms \citep{sterneMultipleImputationMissing2009}.

The most common approach to MI is to specify and fit regression models for partially observed covariates, from which imputations are then generated. Ideally, each one of these imputation models should be compatible with the substantive model of interest. That is, the assumptions made by both models should not conflict with each other, e.g.~the imputation model should at least feature the remaining substantive model covariates, as well as the outcome. We refer to an imputation model as being ``directly specified'' when substantive model covariates and outcome variable(s), or any transformations thereof, are included explicitly as predictors in the imputation model. The use of these directly specified imputation models in settings with missingness spanning multiple covariates is more commonly known as MICE (multivariate imputation by chained equations \cite{buurenMiceMultivariateImputation2011}).

In the context of cause-specific Cox proportional hazards models \citep{prenticeAnalysisFailureTimes1978}, it has been shown that the imputation model for a partially observed covariate should at least include as predictors the other covariates from the substantive model, together with the cause-specific cumulative hazard and event indicator for each competing risk \citep{bonnevilleMultipleImputationCausespecific2022}. Analogously to the standard single-event survival setting, this directly specified imputation model is generally only approximately compatible with the proportional hazards substantive model\citep{whiteImputingMissingCovariate2009}. Concretely, when the outcome model assumes proportional hazards, the conditional distribution of a partially observed covariate modelled using MICE is only an approximation of the ``true'' (i.e.~implied assuming the substantive model is correctly specified) conditional distribution of the partially observed covariate given the outcome and other substantive model covariates. If imputed values can instead be directly sampled from the latter distribution, it would ensure compatibility between analysis and imputation model. This alternative ``indirect'' way of obtaining imputations is referred to as the substantive-model-compatible imputation (SMC-FCS\cite{bartlettMultipleImputationCovariates2015}) approach, and it was extended by Bartlett and Taylor to accommodate cause-specific Cox substantive models\cite{bartlettMissingCovariatesCompeting2016}. In terms of estimating cause-specific hazard ratios, simulation studies have shown that the SMC-FCS approach tends to outperform MICE in cases when the substantive model is correctly specified \citep{bonnevilleMultipleImputationCausespecific2022,bartlettMissingCovariatesCompeting2016}.

When a Fine--Gray subdistribution hazard model \citep{fineProportionalHazardsModel1999} is the substantive model of interest, there has to our knowledge been no research on how one should specify an imputation model for a missing covariate \citep{lauMissingnessSettingCompeting2018}. Nevertheless, MICE is still being used in the presence of missing covariates when the substantive model is a Fine--Gray model, particularly in the context of prediction models. While the structure of the imputation model is rarely reported, articles which do describe their imputation procedure appear to use different approaches. For example, in the prognostic Fine--Gray model presented by Archer et al. (where the primary outcome was time to serious fall resulting in hospital admission or death, with competing death due to other causes), the imputation model for a missing covariate contained the other substantive model covariates, and the cause-specific cumulative hazard and event indicator for each competing risk\cite{archerDevelopmentExternalValidation2022}. In contrast, the MICE procedure reported as part of the prognostic models presented by Clift et al. used cumulative subdistribution hazards in the imputation model\cite{cliftLivingRiskPrediction2020}. Heuristically, it would seem the latter approach is more consistent with the substantive model, as the former imputes approximately compatibly with a cause-specific Cox model structure rather than the Fine--Gray model structure.

In this work, we extend the SMC-FCS approach for missing covariates to accommodate a Fine--Gray substantive model for one of the competing events. In the presence of random right censoring, the core idea is to multiply impute the potential censoring times for individuals failing from competing events in a first step \citep{ruanAnalysesCumulativeIncidence2008}, and thereafter use existing SMC-FCS methodology (originally developed for the standard Cox model\cite{bartlettMultipleImputationCovariates2015}) to impute the missing covariates in a second step.

The structure of the manuscript is as follows. We introduce competing risks notation in Section \ref{sec:notation}. In Section \ref{sec:methods} we outline the proposed method, and thereafter assess its performance in a simulation study in Section \ref{sec:sim_study}. We also provide an illustrative analysis using a dataset from the field of allogeneic hematopoietic stem cell transplantation (alloHCT) in Section \ref{sec:polverelli}. Finally, findings are discussed in Section \ref{sec:discussion}, together with recommendations on how to impute covariates in competing risks settings more generally.

\section{Notation} \label{sec:notation}

We consider a setting in which individuals experience only one of $K$ competing events. We denote the event time as $\tilde{T}$, and the competing event indicator as $\tilde{D} \in \{1,...,K\}$. In practice, individuals are subject to some right-censoring time $C$, meaning we only observe realisations $(t_i, d_i)$ of $T = \min(C,\tilde{T})$ and $D = I(\tilde{T} \leq C)\tilde{D}$, where $I(\cdot)$ is the indicator function and $D = 0$ indicates a right-censored observation. The cause-specific hazard for the $k\textsuperscript{th}$ event is defined as  
\begin{equation*} 
	h_k(t) = \lim_{\Delta t \downarrow 0} \frac{P(t \leq \tilde{T} < t + \Delta t, \tilde{D} = k \given \tilde{T} \geq t)}{\Delta t}.
\end{equation*}
These hazards together make up the event-free survival function,
\begin{equation*}
	P(\tilde{T} > t) = \exp \left\{ - \sum_{k = 1}^{K} \int_{0}^{t} h_k(u)\diff u \right\} = \exp \left\{ - \sum_{k = 1}^{K} H_k(t) \right\},
\end{equation*}
assuming the distribution of $T$ is continuous, and $H_k(t)$ is the cause-specific cumulative hazard function for the $k\textsuperscript{th}$ event. The cause-specific cumulative incidence function is then defined as
\begin{equation*}
	F_k(t) = P(\tilde{T} \leq t, \tilde{D} = k) = \int_{0}^{t}h_k(u)S(u-)\diff u,
\end{equation*}
where $S(u-)$ is the event-free survival probability just prior to $u$. 

The subdistribution hazard for the $k\textsuperscript{th}$ event is defined as
\begin{align*}
	\lambda_k(t) &= \frac{-\diff \log \{1 - F_k(t)\}}{\diff t}, \\
	&= \frac{\diff F_k(t)}{\diff t} \times \{1 - F_k(t)\}^{-1}, 
\end{align*}
which can be thought of as the hazard for the improper random variable $\tilde{V}_k = I(\tilde{D} = k) \times \tilde{T} + I(\tilde{D} \neq k) \times \infty$, for which we can write $F_k(t) = P(\tilde{V}_k \leq t)$ \citep{beyersmannCompetingRisksMultistate2012}. The probability mass at infinity makes $\tilde{V}_k$ improper, i.e.~that its density function does not integrate to one.

Suppose interest lies in modelling the cumulative incidence of one of the competing events, say $D = 1$, conditional on (time-fixed) covariates $\mathbf{Z}$. The Fine--Gray model for cause 1 can be written as 
\begin{equation*}
	\lambda_1(t \given \mathbf{Z}) = \lambda_{0}(t)\exp(\boldsymbol{\beta}^\intercal \mathbf{Z}),
\end{equation*}
with $\lambda_{0}(t)$ being the subdistribution baseline hazard function and $\boldsymbol{\beta}$ representing the effects of covariates $\mathbf{Z}$ on the subdistribution hazard. The cumulative incidence function for cause 1 can then be written as
\begin{equation*}
	F_1(t \given \mathbf{Z}) = 1 - \exp \Biggl\{ -\exp(\boldsymbol{\beta}^\intercal \mathbf{Z}) \int_{0}^{t} \lambda_{0}(u)\diff u \Biggr\},
\end{equation*}
where $\int_{0}^{t} \lambda_{0}(u)\diff u = \Lambda_0(t)$ is the cumulative baseline subdistribution hazard. If we define a baseline cumulative incidence function $F_{0}(t) = 1 - \exp\{-\Lambda_0(t)\}$ (i.e.~the cumulative incidence when $\mathbf{Z} = 0$), the model can also be written as
\begin{equation}
	\label{eq:cuminc_comp}
	F_1(t \given \mathbf{Z}) = 1 - \{1 - F_{0}(t)\}^{\exp(\boldsymbol{\beta}^\intercal \mathbf{Z})}.
\end{equation}
In the presence of random right censoring, the Fine--Gray model is usually fitted by maximising a partial likelihood that uses time-dependent inverse probability of censoring weights (IPCW) \citep{fineProportionalHazardsModel1999}.

\section{MI approaches with a Fine--Gray substantive model}
\label{sec:methods}

We consider a setting with $p$ partially observed covariates $X = X_1,...,X_p$, $q$ fully observed covariates $Z = Z_1,...,Z_q$, and $K = 2$ competing events. We assume that (possibly conditional on $Z$) censoring is independent of both $X$ and the competing risks outcomes $\tilde{T}, \tilde{D}$. We furthermore let $X^{\text{obs}}$ and $X^{\text{mis}}$ respectively denote the observed and missing components of $X$ for an individual, and let $R$ be the vector of observation indicators (equal to 1 if the corresponding element of $X$ is observed, or equal to 0 if it is missing). 

The substantive model of interest is $\lambda_1(t \given X, Z) = \lambda_{0}(t)\exp\{g(X,Z;\boldsymbol{\beta})\}$, which is a Fine--Gray model for cause 1, and where $g(X,Z;\boldsymbol{\beta})$ is a function of $X$ and $Z$, parametrised by $\boldsymbol{\beta}$. In this section, we provide an overview of possible approaches for imputing each partially observed $X_j$. In addition to an approach which imputes compatibly with the assumed substantive model, we also consider alternative methods which are either a) only approximately compatible with the substantive model; or b) impute assuming a different underlying competing risks structure (i.e.~cause-specific proportional hazards). We require that the proposed approaches be valid under the missing-at-random (MAR) assumption, that is, $P(R \given T, D, X, Z) = P(R \given T, D, X^{\text{obs}}, Z)$. 

\subsection{MI based on cause-specific hazards models}

\subsubsection{CS-SMC}
\label{ssec:cs-smc}

A first MI approach to consider is to impute compatibly with cause-specific Cox models, despite the substantive model of interest being a Fine--Gray model for cause 1. As described by Bartlett and Taylor\cite{bartlettMissingCovariatesCompeting2016}, this method relies on the substantive-model-compatible imputation density for $X_j$, given by
\begin{equation}
	f(X_j \given T, D, X_{-j}, Z) \propto f(T, D \given X, Z)f(X_j \given X_{-j},Z), \label{eq:imput_dens}
\end{equation}
where $X_{-j}$ refers to the components of $X$ after removing $X_j$, and $f(\cdot)$ is a density function. For example, $f(T, D \given X, Z)$ is used as shorthand notation for $f_{T, D \given X, Z}(t, d \given x,z)$, that is, the density function for the conditional distribution $T, D \given X, Z$, evaluated at $(t,d)$ for given values $x$ and $z$. 

\sloppy In practice, the substantive model $f(T, D \given X, Z;\psi)$ assumed for $f(T, D \given X, Z)$ is a cause-specific Cox model (one for each competing risk). Therefore, $\psi$ ($\psi \in \Psi$) contains the cumulative baseline hazards and log hazard ratios for each cause-specific hazard. A model $f(X_j \given X_{-j},Z;\phi)$ indexed by $\phi$ ($\phi \in \Phi$), is also assumed for $f(X_j \given X_{-j},Z)$. The idea is then to sample candidate imputed values for the missing $X_j$ using $f(X_j \given X_{-j},Z;\phi)$, and accept these if they also represent draws from a density proportional to  $f(T, D \given X, Z;\psi)f(X_j \given X_{-j},Z;\phi)$. We refer to this method as the cause-specific SMC-FCS approach (CS-SMC).

\subsubsection{CS-Approx}
\label{ssec:cs-approx}

The approximately compatible analogue to the cause-specific SMC-FCS approach is described by Bonneville et al. \cite{bonnevilleMultipleImputationCausespecific2022}. As briefly described in the introduction, this approach involves directly specifying an imputation model $f(X_j \given T, D, X_{-j}, Z;\alpha)$ for $f(X_j \given T, D, X_{-j}, Z)$. In order to ensure approximate compatibility with assumed cause-specific Cox substantive models, the imputation model should include as predictors $X_{-j}$, $Z$, $D$ (as a factor variable), and the (marginal, as obtained using the Nelson--Aalen estimator) cause-specific cumulative hazard for each cause $\hat{H}_k(T)$, evaluated at an individual's event or censoring time. We refer to this method as approximately compatible cause-specific MICE (CS-Approx).

\subsubsection{MI based on the relation between the cause-specific and subdistribution hazards}

The imputations generated by the CS-SMC and CS-Approx approaches will typically not be consistent with the assumption of proportional subdistribution hazards for cause 1 made by the substantive model of interest. This is because, for cause 1, proportionality on the cause-specific hazard scale will generally imply non-proportionality on the subdistribution hazard scale \citep{beyersmannCompetingRisksMultistate2012}. One can derive the functional form of these time-varying covariate effects on the subdistribution hazard scale by using the relation between the subdistribution hazard and the cause-specific hazards\cite{putterRelationCausespecificHazard2020a}. The CS-SMC and CS-Approx approaches can therefore be thought of as procedures to impute (approximately) compatibly with a Fine--Gray model with time-varying covariate effects, the functional form of which are determined by the assumptions made for the cause-specific Cox models of each competing event.

A relevant question at this point is whether the relation between cause-specific and subdistribution hazards can instead be used as part of a procedure to impute compatibly with proportional subdistribution hazards for cause 1. In order to motivate such a procedure, we first note that the conditional density of the observed outcome given covariates used in Equation \eqref{eq:imput_dens} can be written both in terms of cause-specific hazards, and in terms of the cumulative incidence functions, as
\begin{align}
	f(T, D \mid X, Z) &= \{h_1(T \given X, Z)S(T \given X, Z)\}^{I(D = 1)}\{h_2(T \given X, Z)S(T \given X, Z)\}^{I(D = 2)} \nonumber \\
	&\qquad \times S(T \given X, Z)^{1 - I(D = 1) - I(D = 2)}, \nonumber \\
	&= f_1(T \given X, Z)^{I(D = 1)}f_2(T \given X, Z)^{I(D = 2)} \nonumber \\
	&\qquad \times \{1 - F_1(T \given X,Z) - F_2(T \given X, Z)\}^{1 - I(D = 1) - I(D = 2)}, \label{eq:outcome_dens}
\end{align}
with $f_k(t \given X, Z) = \diff F_k(t \given X, Z) / \diff t$  known as the ``subdensity'' for cause $k$ \citep{grayClassKSampleTests1988}. These subdensities, in turn, can be expressed in terms of the subdistribution hazard, as
\begin{align}
	f_k(t \given X, Z) &= \lambda_k(t \given X, Z)\{1 - F_k(t \given X, Z)\}, \nonumber \\
	&=  \lambda_k(t \given X, Z)\exp\{- \Lambda_k(t \given X, Z)\}. \label{eq:subdens_to_subdist}
\end{align}
Specifying a Fine--Gray model for cause 1 is an assumption regarding only part of Equation \eqref{eq:outcome_dens}, namely for any terms involving $f_1(T \given X, Z)$. The practical implication of this is that Equation \eqref{eq:imput_dens} cannot be used to impute the missing $X_j$ without making assumptions about cause 2. One could for example assume (for imputation purposes) a cause-specific Cox model for cause 2, derive the implied $h_1(t \given X, Z)$ using the relation between the subdistribution hazard and the cause-specific hazards, and then use both cause-specific hazards to evaluate $f(T, D \given X, Z)$ in \eqref{eq:imput_dens}. 

Given that a Fine--Gray model is assumed for cause 1, some computational difficulties can be encountered while making assumptions for cause 2. For example, specifying a Fine--Gray model also for cause 2 in the imputation procedure could result in the total failure probability at an observed event time $F_1(T \given X, Z) + F_2(T \given X, Z)$ exceeding 1, meaning we would not be able to draw imputed values using \eqref{eq:imput_dens} for high-risk individuals\citep{AustinTFP2021}. An additional example concerns the approach described in the previous paragraph, where $h_1(t \given X, Z)$ is derived based on $h_2(t \given X, Z)$ and $\lambda_1(t \given X, Z)$. The numerical integration step generally needed to compute $h_1(t \given X, Z)$ could make the overall imputation procedure rather computationally inefficient. More details on potential issues when specifying a model for cause 2 when a Fine--Gray model is assumed for cause 1 can be found in Bonneville et al. \cite{bonnevilleWhyYouShould2024}. 

The above points mean that it is desirable to use an alternative approach which avoids having to specify a model for the cause-specific (or subdistribution) hazard of cause 2. In the next subsection, we propose a SMC-FCS approach assuming a Fine--Gray substantive model for cause 1, which avoids making explicit modelling assumptions concerning cause 2.

\subsection{MI based on the Fine--Gray model} \label{sec:subdist_time}

\subsubsection{FG-SMC}
\label{ssec:fg-smc}

Suppose for now that the potential censoring time $C$ is known for all individuals. This is for example the case when there is a fixed end of study date (i.e.~``administrative'' censoring), and no additional random right-censoring. Fine and Gray referred to these kind of data as ``censoring complete'', since the subdistribution at-risk process is known\cite{fineProportionalHazardsModel1999}. Equivalently, the ``observed'' subdistribution random variable for cause 1 (henceforth referred to as ``subdistribution time''), $V = I(D = 1) \times T + I(D \neq 1) \times C$, is known for all individuals. In turn, this implies that (with complete covariate data), the Fine--Gray model can be estimated by fitting a standard Cox model with outcome $V$ and event indicator $I(D = 1)$.

Consequently, an intuitive approach to imputing the missing $X_j$ in our setting might therefore be to apply existing SMC-FCS methodology for standard Cox models (see section 6.3 of Bartlett et al.\cite{bartlettMultipleImputationCovariates2015}), but instead using $V$ and $I(D = 1)$ as our outcome variables. We refer to this method as Fine--Gray SMC-FCS (FG-SMC). The substantive-model-compatible imputation density is now
\begin{equation}
	f(X_j \given V, D, Z) \propto f(V, D \given X, Z)f(X_j \given X_{-j},Z), \label{eq:imput_dens_V}
\end{equation}
where the conditional density of the observed outcome given the covariates can be written as
\begin{align}
	f(V, D \mid X, Z) &= f_1(V \mid X, Z)^{I(D = 1)}\{1 - F_1(V \mid X, Z)\}^{I(D \neq 1)}, \nonumber \\
	&= [\lambda_1(V \given X, Z)\exp\{- \Lambda_1(V \given X, Z)\}]^{I(D = 1)}\exp\{- \Lambda_1(V \given X, Z)\}^{I(D = 0)} \nonumber \\
	&\qquad \exp\{- \Lambda_1(V \given X, Z)\}^{I(D = 2)}, \nonumber  \\
	&= \lambda_1(V \given X, Z)^{I(D = 1)}\exp\{- \Lambda_1(V \given X, Z)\}, \label{eq:parallel_cox}
\end{align}
using Equation \eqref{eq:subdens_to_subdist} and the fact that $f_1(V \given X, Z)^{I(D = 1)} = f_1(T \given X, Z)^{I(D = 1)}$. Note that while Equations \eqref{eq:imput_dens_V} and \eqref{eq:parallel_cox} depend only on $I(D = 1)$,  we still use $D$ in the notation to make the contribution of those failing from cause 2 to the density explicit, which is relevant for the upcoming sections. Importantly, this procedure relies on a stronger MAR assumption (compared to the one introduced at the beginning of Section \ref{sec:methods}), namely $P\{R \given V, I(D = 1), X, Z\} = P\{R \given V, I(D = 1), X^{\text{obs}}, Z\}$. In essence, we ignore any terms involving $f_2(T \given X, Z)$ in Equation \eqref{eq:imput_dens} based on the assumption that missingness in $X$ does not depend on either $I(D = 2)$ or the failure time for those failing from cause 2.

\subsubsection{FG-Approx}
\label{ssec:fg-approx}

The form of Equation \eqref{eq:parallel_cox} mirrors the likelihood in the standard Cox context, which can be obtained by replacing $\lambda_1(V \given X, Z)$ with the hazard of a single event (in absence of competing risks). The practical implications of this for our MI context are that the findings of White and Royston\cite{whiteImputingMissingCovariate2009} in the single-event survival setting should in principle extend to the Fine--Gray context. Namely, that the (approximately compatible) directly specified imputation model $f(X_j \given V, D, X_{-j}, Z;\alpha)$ for a partially observed $X_j$ should contain as predictors at least $X_{-j}$, $Z$, the indicator for the competing event of interest $I(D = 1)$, and the cumulative subdistribution baseline hazard for the same event $\Lambda_0(V)$. Instead of the unknown true $\Lambda_0(V)$, one could use the estimated marginal cumulative subdistribution hazard $\hat{\Lambda}_1(V)$ instead, obtained using the Nelson--Aalen estimator using $V$ and $I(D = 1)$ are outcome variables. We refer to this approximately compatible MICE approach as FG-Approx.

\subsection{Accommodating random right-censoring}

In addition to (deterministic) administrative censoring, random right-censoring may occur. In the presence of random right-censoring, the contribution of those failing from cause 2 to density \eqref{eq:parallel_cox} is no longer evaluable, since we do not know their potential censoring time. In a sense, their subdistribution time has been informatively censored by their cause 2 failure. 

\subsubsection{Via imputation of potential censoring times} \label{ssec:imp_cens}

One approach to estimate the parameters of a Fine--Gray model in the presence of random right censoring is to consider the potential censoring times for those failing from cause 2 as missing data, and multiply impute them. To this end, Ruan and Gray\cite{ruanAnalysesCumulativeIncidence2008} suggested the use of Kaplan--Meier (KM) imputation \citep{taylor2002survival}. Specifically, potential censoring times are randomly drawn from the conditional distribution with distribution function $1 - P(C > t \given C > T) = 1 - \hat{G}(t-)/\hat{G}(T-)$, where $\hat{G}(t)$ is a KM estimate of the survival distribution of the censoring times $P(C > t)$. The imputation of these potential censoring times effectively produces multiple ``censoring complete'' datasets, in which a Fine--Gray model can be fit using standard software. Inference is then based on a pooled model, which combines the models fitted in each censoring complete dataset using Rubin's rules \citep{rubin:1987}.

We can make use of the above ideas in order to multiply impute covariates compatibly with a Fine--Gray model in the presence of random right random censoring. Specifically, we can apply the FG-SMC (or FG-Approx) method in each censoring complete dataset obtained after first imputing the potential censoring times for those failing from cause 2. In order to formalise this procedure, recall that $\boldsymbol{\beta}$ represents the parameters of the substantive model, and that $X = \{X^{\text{obs}},X^{\text{mis}}\}$. We can similarly partition $V = \{V^{\text{obs}},V^{\text{mis}}\}$, where $V^{\text{mis}}$ is the vector of missing censoring times for those failing from cause 2. 

From a Bayesian perspective, the goal is to estimate the conditional density of $\boldsymbol{\beta}$ given the observed data, namely
\begin{align}
	f(\boldsymbol{\beta} \given X^{\text{obs}},Z,V^{\text{obs}},D) &= \int_V \int_X f(\boldsymbol{\beta} \given X^{\text{obs}},X^{\text{mis}},Z,V^{\text{obs}},V^{\text{mis}},D) \times \nonumber \\ 
	&\qquad f(X^{\text{mis}}, V^{\text{mis}} \given X^{\text{obs}},Z,V^{\text{obs}},D)\diff X^{\text{mis}}\diff V^{\text{mis}}. \label{eq:posterior_beta}
\end{align} 
If we can sample imputed values $M$ times from $f(X^{\text{mis}}, V^{\text{mis}} \given X^{\text{obs}},Z,V^{\text{obs}},D)$, the integral above can be approximated by an average over $f(\boldsymbol{\beta} \given X^{\text{obs}},X^{\text{mis}},Z,V^{\text{obs}},V^{\text{mis}},D)$ (the ``complete data'' posterior density) evaluated at those $M$ moments \citep{molenberghs2014handbook}.

One option to sample from $f(X^{\text{mis}}, V^{\text{mis}} \given X^{\text{obs}},Z,V^{\text{obs}},D)$, the joint posterior predictive density, is to iteratively sample from $f(V^{\text{mis}} \given X,Z,V^{\text{obs}},D)$ and $f(X^{\text{mis}} \given X^{\text{obs}},Z,V,D)$. This is clearly necessary when the censoring distribution depends on $X$, since the potential censoring times will differ depending on the most recently imputed $X$. These imputed censoring times would have to be based for example on a Cox model for the censoring hazard instead of a marginal KM estimate. 

\begin{figure}[ht]
	\centering
	\includegraphics[width=\textwidth]{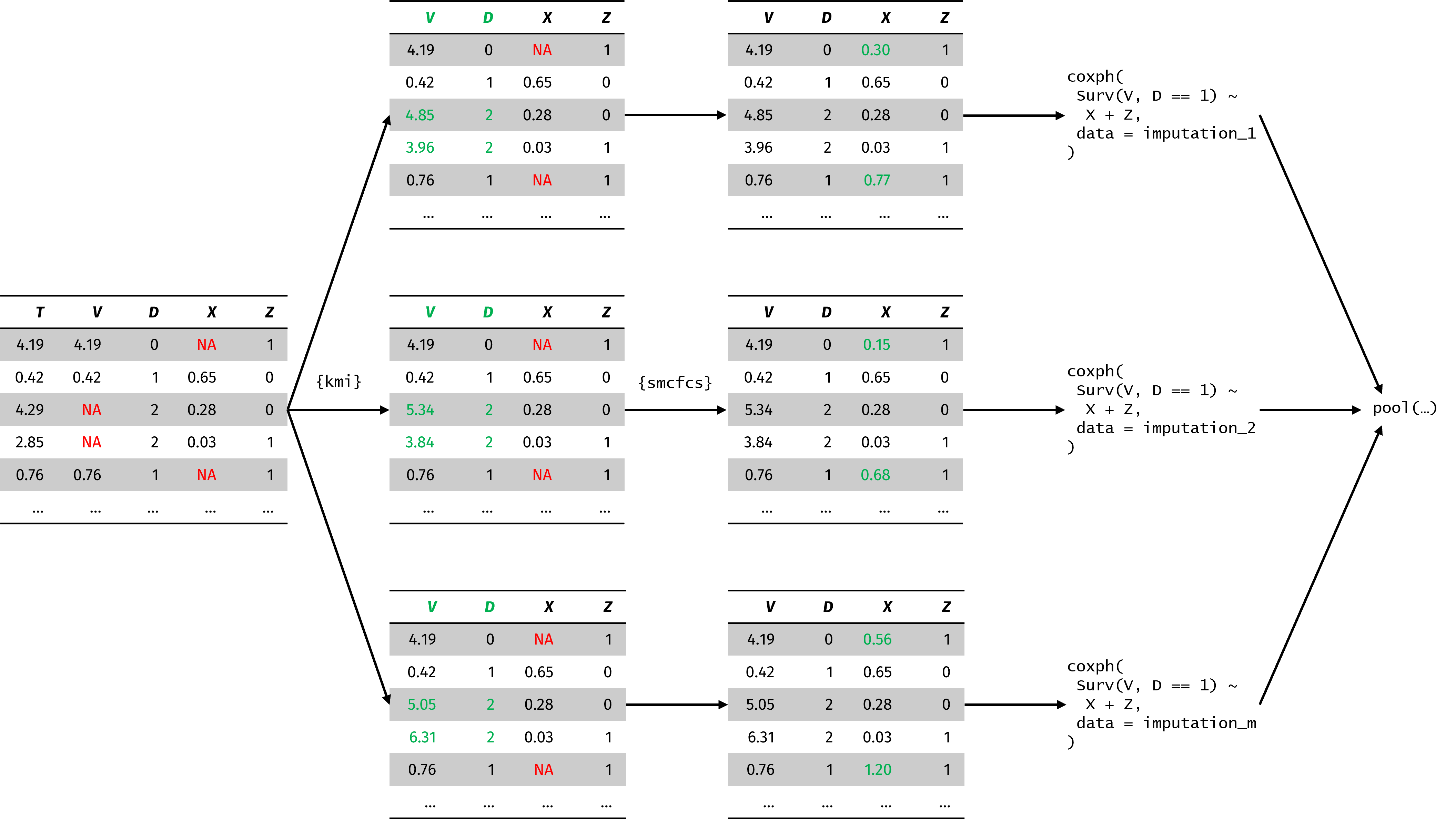}
	\caption{Sequential workflow for (compatible) covariate imputation and analysis for a Fine--Gray substantive model with two covariates $X$ and $Z$, in the presence of random right-censoring. In the first step, the potential censoring times for those failing from cause 2 are multiple imputed using the \{kmi\} package. In the second step, the missing covariates are imputed using the \{smcfcs\} (or \{mice\}) package. This workflow is valid when the probability of being censored is independent of $X$ and any $Z$ related to the censoring process are modelled in \{kmi\}.}\label{fig:workflow}
\end{figure}

If however $C \indep X \given Z$, we could use a sequential approach, where we factorise
\begin{align*}
	f(X^{\text{mis}}, V^{\text{mis}} \given X^{\text{obs}},Z,V^{\text{obs}},D) &= f(X^{\text{mis}} \given X^{\text{obs}},Z,V,D)f(V^{\text{mis}} \given X^{\text{obs}},Z,V^{\text{obs}},D), \\
	&= f(X^{\text{mis}} \given X^{\text{obs}},Z,V,D)f(V^{\text{mis}} \given Z,V^{\text{obs}},D).
\end{align*} 
Practically speaking, this involves imputing the potential censoring times (possibly in strata of $Z$) in a first step, and then imputing the missing $X$ in a second step. This can be implemented easily using existing software packages in R: \{kmi\} for the imputation of censoring times \citep{allignolSoftwareFittingNonstandard2010}, and \{smcfcs\} for the imputation of the missing covariates \citep{bartlettSmcfcsMultipleImputation2022a} - see Figure \ref{fig:workflow} for an illustration of the workflow.

Note that the described KM-based procedure for imputing potential censoring times does not take into account any of the uncertainty in estimating $P(C > t)$. Ruan and Gray discussed using the non-parametric bootstrap to account for this uncertainty and improve estimation properties, and found similar results both with and without a bootstrap step\cite{ruanAnalysesCumulativeIncidence2008}. In Appendix \ref{sec:appendix_subdist}, we visualise and give additional details concerning the imputation of potential censoring times. 

\subsubsection{Via censoring weights in the likelihood}

Rather than multiply imputing the potential censoring times, an alternative approach is to incorporate inverse probability of censoring weights directly in \eqref{eq:parallel_cox}. If we define time-dependent weights
\begin{equation*}
	\begin{aligned}
		w(t) &= 1 && \text{if } t \leq T,\\
		w(t) &= P(C > t \given C > T) = \frac{G(t-)}{G(T-)} && \text{if } t > T,
	\end{aligned}
\end{equation*}
then the conditional density of the (subdistribution) outcome given the covariates can be written as
\begin{align}
	f(V, D \mid X, Z) &= [\lambda_1(V \given X, Z)\exp\{- \Lambda_1(V \given X, Z)\}]^{I(D = 1)}\exp\{- \Lambda_1(V \given X, Z)\}^{I(D = 0)} \times \nonumber \\
	&\qquad \exp\Biggl\{- \int_{0}^{\infty}w(u)\lambda_1(u\given X, Z)\diff u\Biggr\}^{I(D = 2)}, \label{eq:weighted_lik}
\end{align}
where the term for those failing from the competing event involves integration in practice up to a maximum potential follow-up time $t^*$. As described by Lambert et al., this integral can be approximated by splitting time into intervals, in which the corresponding $w(t)$ is assumed to be constant\cite{lambertFlexibleParametricModelling2017}.

The integration step needed for those failing from cause 2 in \eqref{eq:weighted_lik} means that this approach cannot be implemented in a straightforward way with existing software, unlike the approach described in the previous subsection. The simulation study in this paper therefore focuses on the approach involving multiple imputation of potential censoring times.

\subsection{Implementation of MI approaches}

Methods CS-SMC, CS-Approx, FG-SMC, and FG-Approx can all be implemented using existing software packages in R. In this section, we summarise the steps needed to apply these methods in a given dataset in the presence of random right censoring (possibly in combination with administrative censoring). A minimal R code example can be found in supplementary material S1.

\begin{enumerate}
	\item Add columns $\hat{H}_1(T)$ and $\hat{H}_2(T)$ to the original data, which are the marginal cause-specific cumulative hazards for each competing risk evaluated at an individual's event or censoring time (obtained using the Nelson--Aalen estimator).
	\item Multiply impute the potential censoring for those failing from cause 2 using \{kmi\}, yielding $m$ censoring complete datasets (i.e.~with ``complete'' $V$). The censoring distribution has support at both random and administrative censoring times. Any completely observed covariates that are known to affect the probability of being censored should be included as predictors in the model for the censoring process. \{kmi\} imputes based on stratified KM when $Z$ are categorical, and based on a Cox model at least one of $Z$ is continuous. If for example an individual's time of entry into a study determines their maximum follow-up duration, this should be accounted for in the imputation procedure (e.g.~by stratifying by year of entry).
	\item In each censoring complete dataset, add an additional column $\hat{\Lambda}_1(V)$. This takes the value of the marginal cumulative subdistribution hazard for cause 1 at an individual's observed or imputed subdistribution time, obtained with the Nelson--Aalen estimator based on $I(D = 1)$ and imputed $V$.
	\item In each censoring complete dataset (each with different $V$ and $\hat{\Lambda}_1(V)$, but same $\hat{H}_1(T)$ and $\hat{H}_2(T)$), create a single imputed dataset using the desired covariate imputation method(s):
	\begin{itemize}
		\item CS-SMC: use \{smcfcs\} to impute the missing covariate(s) compatibly with cause-specific Cox models. All covariates used in the Fine--Gray substantive model should feature in at least one of the specified cause-specific models.
		\item CS-Approx: use \{mice\} to impute the missing covariate(s), where the imputation model contains as predictors the remaining substantive model covariates, $D$ (as a factor variable), and both $\hat{H}_1(T)$ and $\hat{H}_2(T)$.
		\item FG-SMC: use \{smcfcs\} to impute the missing covariate(s) compatibly with the Fine--Gray substantive model. This is done by using the imputation methods developed for the standard Cox model, but with as outcome variables $I(D = 1)$ and imputed $V$.
		\item FG-Approx: use \{mice\} to impute the missing covariate(s), where the imputation model contains as predictors the remaining substantive model covariates, $I(D = 1)$, and $\hat{\Lambda}_1(V)$.
	\end{itemize}
	\item Fit the Fine--Gray substantive model in each imputed dataset (using standard Cox software with $I(D = 1)$ and imputed $V$ as outcome variables), and pool the estimates using Rubin's rules.
\end{enumerate}

\section{Simulation study}
\label{sec:sim_study}

We aim to evaluate the performance of different MI methods in the presence of missing covariate data when specifying a Fine--Gray model for the subdistribution hazard for one event of interest in the presence of one competing event. Specifically, we assume interest lies in the estimation (for cause 1) of both subdistribution hazard ratios, and the cumulative incidence for a particular individual at some future time horizon. We follow the ADEMP structure for the reporting of the simulation study\cite{morrisUsingSimulationStudies2019a}.

\subsection{Data-generating mechanisms}

We generate datasets of $n = 2000$ individuals, with two covariates $X$ and $Z$. We assume $Z \sim \mathcal{N}(0,1)$ and $X \given Z \sim \text{Bernoulli}\{(1 + e^{-Z})^{-1}\}$. 

We let $h_k(t \given X, Z)$,  $\lambda_k(t \given X, Z)$ and $F_k(t \given X, Z) = P(\tilde{T} \leq t, \tilde{D} = k \given X, Z)$ respectively denote the cause-specific hazards, subdistribution hazards and cumulative incidence functions for cause $k$, conditional on $X$ and $Z$. The competing event times will be generated following two mechanisms: one where the Fine--Gray model for cause 1 is correctly specified, and another where it is misspecified. These are detailed below, together with assumptions concerning both censoring and the missing data mechanisms.

\subsubsection{Correctly specified Fine--Gray} \label{corr-spec-FG}

For this mechanism, we simulate data using the `indirect' method described in Beyersmann et al.\cite{beyersmannCompetingRisksMultistate2012}, and originally used in the simulations by Fine and Gray \cite{fineProportionalHazardsModel1999}. This approach involves first drawing the competing event indicator $\tilde{D}$, and then generating an event time for those with $\tilde{D} = 1$. The final step is to generate times of the competing event for the remaining individuals, who were assigned $\tilde{D} = 2$. 

Here, we directly specify the cumulative incidence of cause 1 as
\begin{equation*}
	F_1(t \given X, Z) = 1 - \big[1 - p\{1- \exp(-b_1t^{a_1})\}\big]^{\exp(\beta_{1}X + \beta_{2}Z)}.
\end{equation*}
The above expression corresponds to a Fine--Gray model, with as baseline cumulative incidence function a Weibull cumulative distribution function with shape $a_1$ and rate $b_1$ (parametrisation used in Klein and Moeschberger\cite{kleinSurvivalAnalysisTechniques2006}) multiplied by a probability $p$. Explicitly,
\begin{equation*}
	F_0(t) = p\{1- \exp(-b_1t^{a_1})\}.
\end{equation*}
With $\lim_{t \to \infty}F_0(t) = p$, we have that 
$P(\tilde{D} = 1 \given X,Z) = 1 - (1-p)^{\exp(\beta_{1}X + \beta_{2}Z)}$, and $P(\tilde{D} = 2 \given X, Z) = 1 - P(\tilde{D} = 1 \given X, Z) = (1-p)^{\exp(\beta_{1}X + \beta_{2}Z)}$. These are the individual-specific cumulative incidences for each event at time infinity. Also note that the baseline subdistribution hazard for this mechanism can be obtained by $\{\diff F_0(t) / \diff t\} \times \{1 - F_0(t)\}^{-1}$.

The idea then is to generate the event times for cause 1 conditionally on the event indicator and covariates, using 
\begin{align}
	P(\tilde{T} \leq t \given \tilde{D}=1, X, Z) &= \frac{P(\tilde{T} \leq t, \tilde{D}=1 \given X, Z)}{P(\tilde{D} = 1 \given X, Z)} \nonumber \\
	&= \frac{1 - \big[1-p\{1- \exp(-b_1t^{a_1})\}\big]^{\exp(\beta_{1}X + \beta_{2}Z)}}{1 - (1-p)^{\exp(\beta_{1}X + \beta_{2}Z)}}. \label{eq:expr_to_invert}
\end{align}
To sample from the above, we first need to draw $\tilde{D} \sim \text{Bernoulli}\{(1-p)^{\exp(\beta_{1}X + \beta_{2}Z)}\} + 1$. We can then use inverse transform sampling to draw failure times within the subset of individuals with $\tilde{D} = 1$. Shortening $\exp(\beta_{1}X + \beta_{2}Z) = \exp(\eta)$, and with $u \sim \mathcal{U}(0,1)$, we can invert \eqref{eq:expr_to_invert} as
\begin{equation*}
	t = \Biggl[- \frac{1}{b_1}\log\Bigg[1 - \frac{1-\big[ 1 - u\{1 - (1-p)^{\exp(\eta)}\} \big]^{1/\exp(\eta)}}{p} \Bigg]\Biggr]^{1/{a_1}}.
\end{equation*}
For the competing event, we can factorise the cumulative incidence function as
\begin{equation*}
	P(\tilde{T} \leq t, D=2 \given X, Z) = P(\tilde{T} \leq t \given \tilde{D}=2,  X, Z)P(\tilde{D}=2 \given X, Z).
\end{equation*}
A proportional hazards model can then be specified (for convenience) for 
\begin{equation*}
	P(\tilde{T} \leq t \given \tilde{D}=2,  X, Z) = 1 - \exp\Bigl\{-H_{20}^*(t)\exp(\beta_{1}^*X + \beta_{2}^*Z) \Bigr\},
\end{equation*}
where $H_{20}^*(t)$ is the cumulative baseline hazard associated to the cumulative incidence function conditional on $\tilde{D} = 2$. Since the event indicator is already drawn, the failure times can be drawn using standard methods within the subset with $\tilde{D} = 2$. Here, we specify a Weibull baseline hazard as $h_{20}^*(t) = a_2b_2t^{a_2 - 1}$.

We fix $\{\beta_{1}, \beta_{2},\beta_{1}^*,\beta_{2}^*\} = \{0.75, 0.5, 0.75, 0.5\}$, and the Weibull parameters used for both events as shape $\{a_1,a_2\} = 0.75$ and rate $\{b_1,b_2\} = 1$. We vary $p = \{0.15, 0.65\}$, which is the expected proportion of event 1 failures for individuals with $X = 0$ and $Z = 0$.

\subsubsection{Simulation based on cause-specific hazards (misspecified Fine--Gray)}

In this data-generating mechanism (DGM), we assume proportionality on the cause-specific hazard scale, and simulate using latent failure times \citep{beyersmannSimulatingCompetingRisks2009}. We specify baseline Weibull hazards for both cause-specific hazards as 
\begin{align*}
	h_1(t \given X, Z) &= a_1b_1t^{a_1 - 1}\exp(\gamma_{11}X + \gamma_{12}Z), \\
	h_2(t \given X, Z) &= a_2b_2t^{a_2 - 1}\exp(\gamma_{21}X + \gamma_{22}Z),
\end{align*}
where $\{a_1,a_2\}$ and $\{b_1,b_2\}$ are respectively the shape and rate parameters. Under this DGM, a Fine--Gray model for cause 1 will be misspecified. Nevertheless, the coefficients resulting from the misspecified Fine--Gray model could still be interpreted as time-averaged effects on the (complementary log-log transformed) cumulative incidence function\cite{grambauerProportionalSubdistributionHazards2010a}. 

We aim to have a scenario close to the one described in \ref{corr-spec-FG} (in terms of event proportions), where the main difference is that proportionality now holds on the cause-specific hazard scale. To fix the parameters in this DGM, we first simulate a large dataset of one million individuals following the mechanism described in the previous subsection, where proportional subdistribution hazards hold. Parametric cause-specific proportional hazards models assuming baseline Weibull hazards are then fitted for each failure cause. The point estimates obtained from these models are used as the cause-specific data-generating parameters $\{a_1,b_1, \gamma_{11}, \gamma_{12}\}$ and $\{a_2,b_2, \gamma_{21}, \gamma_{22}\}$. These parameters will of course differ depending on $p = \{0.15, 0.65\}$, and also depending on the censoring distribution. While the cause-specific models fitted on this large dataset will be misspecified (cause-specific baseline hazards are not of Weibull shape, and covariates effects on the cause-specific hazards are non-proportional), the resulting ``least false'' parameters are still useful. 

\begin{figure}[ht]
	\centering
	\includegraphics[width=\textwidth]{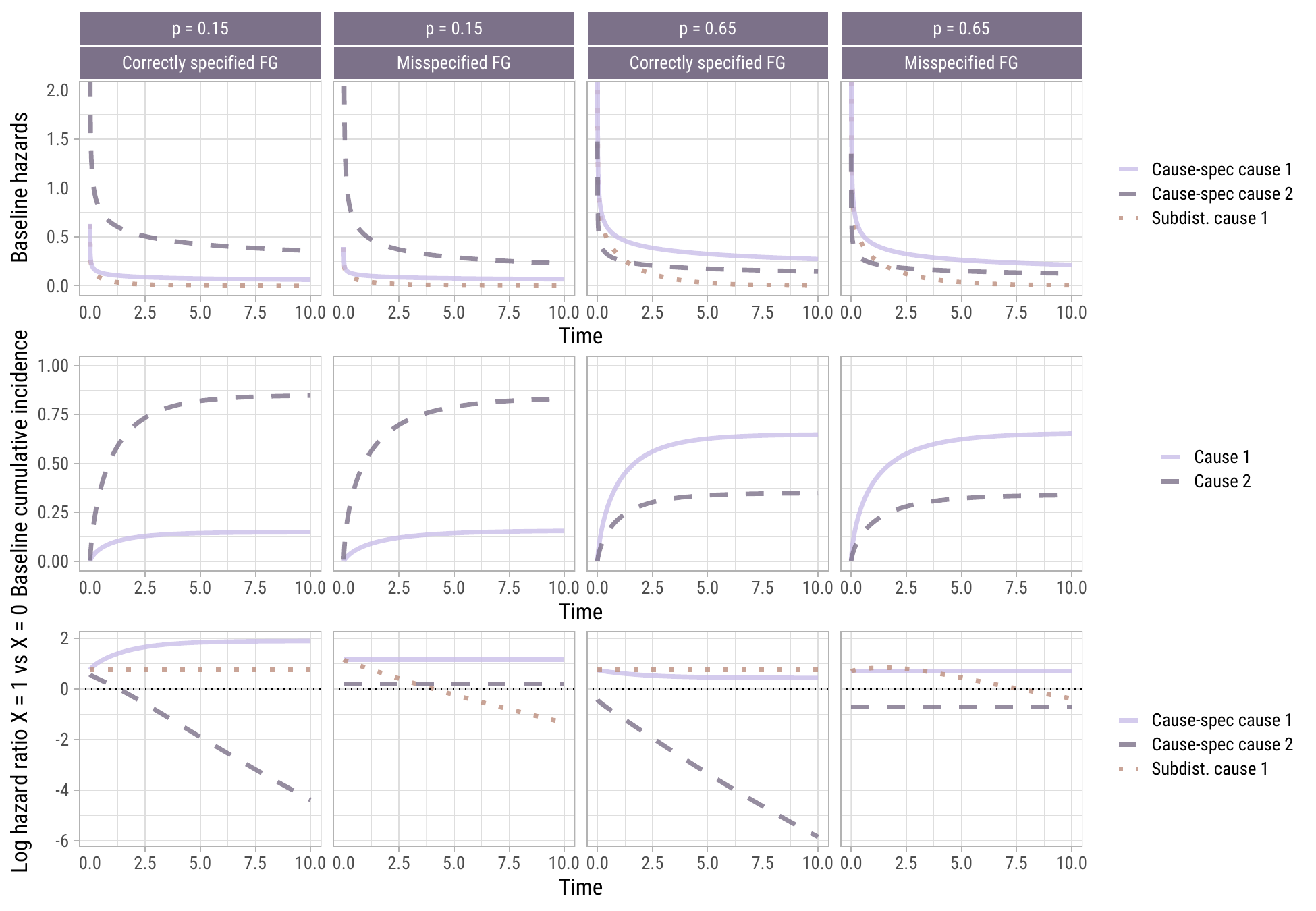}
	\caption{Summary of data-generating mechanisms prior to the addition of any censoring. For each value of $p$, both the correctly specified and misspecified Fine--Gray (FG) scenarios are very comparable in terms of (true) baseline hazards and cumulative incidences.}
	\label{fig:scenarios_vis}
\end{figure}

Figure \ref{fig:scenarios_vis} summarises the DGMs, prior to the addition of any censoring. In the correctly specified Fine--Gray scenarios, the subdistribution log hazard ratio $\lambda_1(t \given X = 1, Z)/\lambda_1(t \given X = 0, Z)$ is time constant, while the cause-specific log hazard ratios are time-dependent. The reverse is true for the misspecified Fine--Gray scenarios. Overall, the correctly specified and misspecified Fine--Gray scenarios are very comparable in terms of (true) baseline hazards and cumulative incidences, for both values of $p$.

\subsubsection{Censoring}

The DGMs outlined above assume no loss to follow-up. As additional scenarios, we consider independent (i.e.~not conditional on any covariates) right censoring where the censoring times are simulated from an exponential distribution with rate $\lambda_C = 0.49$, resulting in approximately 30\% of censored observations. These censoring times will be considered as either: a) known (administrative censoring); or b) unknown (random censoring).

\subsubsection{Covariate missingness}

Missingness is induced in $X$, while $Z$ remains fully observed. Let $R_X$ be a binary variable indicating whether $X$ is missing ($R_X = 0$) or observed ($R_X = 1$). We use a missing at random (MAR) mechanism conditional on $Z$, which was defined as $\logit P(R_X = 0 \given Z) = \eta_0 + \eta_1 Z$. We take $\eta_1 = 1.5$, a rather strong mechanism where higher values of $Z$ are associated with more missingness in $X$. The value of $\eta_0$ is found via standard root solving, such that the average probability $P(R_X = 0) = \mathbb{E}\{P(R_X = 0 \given Z)\}$ of being missing in a given dataset equals 0.4. 

\subsubsection{Summary}

In summary, the simulation study varied

\begin{itemize}
	\item Censoring type: no censoring, administrative and random censoring
	\item Relative occurrence of event 1, as low or high. This is done by varying the baseline cumulative incidence of event 1 (as $t \rightarrow \infty$) as $p = \{0.15, 0.65\}$.
	\item Failure time simulation methods, with a) directly specified cumulative incidence cause 1 (correctly specified Fine--Gray); b) cause-specific proportional hazards for both causes (misspecified Fine--Gray).
\end{itemize}

This adds up to 3 (censoring types) $\times$ 2 (relative occurence event 1) $\times$ 2 (failure time simulation methods) = 12 scenarios.

\subsection{Estimands}

The first estimands of interest are the subdistribution log hazard ratios $\beta_{1}$ and $\beta_{2}$ for $X$ and $Z$, respectively. In the correctly specified Fine--Gray scenarios, these simply correspond to the data-generating parameters $\{\beta_{1},\beta_{2}\} = \{0.75, 0.5\}$. In the misspecified Fine--Gray scenarios however, the target values (the ``least-false parameters''; time averaged subdistribution log hazard ratios $\{\tilde{\beta}_{1},\tilde{\beta}_{2}\}$) are obtained by fitting a Fine--Gray model on a large simulated dataset of one million individuals, simulated as under the second data-generating mechanism, after applying any censoring. For computational efficiency, the censoring times are assumed to be known when fitting the Fine--Gray model on this large dataset.

The second estimands of interest are the conditional cumulative incidence of event 1 at a grid of timepoints (between timepoints 0 and 5) for reference individuals $\{X, Z\} = \{0,0\}$ (baseline) and $\{X, Z\} = \{1,1\}$. In the correctly specified Fine--Gray scenarios, this corresponds to
\begin{equation*}
	F_1(t \given X, Z) = 1 - \big[1-p\{1- \exp(-b_1t^{a_1})\}\big]^{\exp(\beta_{1}X + \beta_{2}Z)},
\end{equation*}
while for the misspecified Fine--Gray scenarios, this corresponds to
\begin{align*}
	F_1(t \given X, Z) &= \int_{0}^{t}h_1(u \given X, Z) \exp\Bigl\{{ - H_1(u \given X, Z) - H_2(u \given X, Z) \Bigr\}}\diff u, \\
	&= \int_{0}^{t} a_1b_1u^{a_1 - 1}\exp(\gamma_{11}X + \gamma_{12}Z) \\
	&\qquad \times  \exp\Bigl\{{ -b_1u^{a_1}\exp(\gamma_{11}X + \gamma_{12}Z) - b_2u^{a_2}\exp(\gamma_{21}X + \gamma_{22}Z) \Bigr\}}\diff u,
\end{align*}
which is obtained via numerical integration.

\subsection{Methods}

The assessed methods are

\begin{itemize}
	\item Full: analysis run on full data prior to missing values, as a benchmark for the best possible performance.
	\item CCA: complete-case analysis, as a ``lower'' benchmark that the imputation methods need to outperform in order to be worthwhile.
	\item CS-SMC: MI, imputing compatibly with cause-specific Cox proportional hazards models. This method is described in Section \ref{ssec:cs-smc}. Both $X$ and $Z$ are used as predictors in each cause-specific model assumed by this procedure. 
	\item CS-Approx: MI with both marginal cumulative cause-specific hazards (evaluated at the individual observed event or censoring time) and competing event indicator included as predictors in the imputation model, in addition to $Z$. This method is described in Section \ref{ssec:cs-approx}.
	\item FG-SMC: MI, imputing compatibly with a Fine--Gray model for cause 1 that has as covariates $X$ and $Z$. This is the method described in Section \ref{ssec:fg-smc}.
	\item FG-Approx: MI with marginal cumulative subdistribution hazard (evaluated at the individual observed or imputed subdistribution time $V$) and indicator for event 1 included as predictors in the imputation model, in addition to $Z$. This method is described in Section \ref{ssec:fg-approx}. 
\end{itemize}

The imputation methods are run with 30 imputed datasets. This was fixed following a pilot set of simulations with 50 imputed datasets, which showed that there was little reduction in empirical standard errors for the subdistribution log hazard ratios (and their Monte Carlo standard errors) beyond 30 imputed datasets. Approximately compatible MI methods CS-Approx and FG-Approx only require a single iteration because there is just one variable with missing values, while substantive-model-compatible (SMC) MI methods CS-SMC and FG-SMC are run with 20 iterations. The method used to model $f(X \given V, D, Z;\alpha)$ for approximately compatible methods is logistic regression, while for SMC methods $f(X \given Z;\psi)$ is specified as a logistic regression. We note that $X$ was chosen to be binary as SMC methods do not require rejection sampling for variables with discrete sample space, thereby reducing simulation time.

For the scenarios with no or administrative censoring, the subdistribution time $V$ is fully observed. While $V = T$ for those failing from cause 1, for those failing from cause 2, $V$ is first set to either a) a large value greater than the largest observed event 1 time (in absence of censoring); or b) the known potential censoring time $C$ (administrative censoring). The marginal cumulative subdistribution hazard used for the approximate subdistribution MI method is obtained using a marginal model with $I(D=1)$ and the resulting $V$ as outcome variables. The covariate MI methods are run once these $V$ and $I(D=1)$ variables have been created. In scenarios with random censoring, the potential censoring times for those failing from cause 2 are multiply imputed using the \{kmi\} R package with default settings: marginal non-parametric model for the censoring distribution, and no additional bootstrap layer. This yields 30 imputed datasets, each with a different $V$. In each of these datasets, the marginal cumulative subdistribution hazard is estimated in the same way as described above. Thereafter, the covariate MI methods are run in each of these datasets, yielding one imputed dataset for each imputed $V$ (total of 30 imputed datasets), corresponding to the workflow in Figure \ref{fig:workflow}.

For all methods, the Fine--Gray model for cause 1 is estimated using a Cox model with (known or imputed) $V$ and $I(D = 1)$ as outcome variables. When the imputation methods are used (and for all methods when there is random right censoring), the estimated $\hat{\beta}_{1}$ and $\hat{\beta}_{2}$ are the results of coefficients pooled using Rubin's rules. Confidence intervals around these estimates are built as described in Section 2.4.2 in the text by van Buuren \cite{buurenFlexibleImputationMissing2018}. For the cumulative incidences, the estimates for the two sets of reference values of $X$ and $Z$ are first made in \textit{each} imputed dataset using Equation \eqref{eq:cuminc_comp}, and thereafter pooled using Rubin's rules after complementary log-log transformation - as described in Morisot et al.\cite{morisotProstateCancerNet2015} and recommended by Marshall et al.\cite{marshallCombiningEstimatesInterest2009}. This predict-then-pool approach (rather than predicting using a pooled model) has been recommended by multiple authors \cite{woodEstimationUsePredictions2015,mertensConstructionAssessmentPrediction2020}.

\subsubsection{Performance measures}

The primary measure of interest was bias in the estimated subdistribution log hazard ratios. In order to keep the Monte Carlo standard error (MCSE) of bias under a desired threshold of 0.01, we require $n_{\text{sim}} = 0.2^2/0.01^2 = 400$ replications per scenario, as we expect empirical standard errors to be under 0.2 for all scenarios (based on a pilot run). This number was rounded up to $n_{\text{sim}} = 500$. In addition to bias, we recorded empirical and estimated standard errors, and coverage probabilities. For the cumulative incidence estimates, we focused on both bias and root mean square error (RMSE).

\subsubsection{Software}

Analyses were performed using R version 4.3.1 \citep{rcoreteamLanguageEnvironmentStatistical2023}. Core packages used were: \{survival\} version 3.5.7 \citep{survival-package}, \{mice\} version 3.16.0 \citep{buurenMiceMultivariateImputation2011}, \{smcfcs\} version 1.7.1 \citep{bartlettSmcfcsMultipleImputation2022a}, \{kmi\} version 0.5.5 \citep{allignolSoftwareFittingNonstandard2010}, and \{rsimsum\} version 0.11.3 \citep{gaspariniRsimsumSummariseResults2018}.

\subsection{Results}

We summarise the main findings in this section, with full results available in a markdown file on the Github repository linked at the end of the present manuscript.


\subsubsection{Subdistribution log hazard ratios}

\begin{figure}[ht]
	\includegraphics[width=0.85\textwidth]{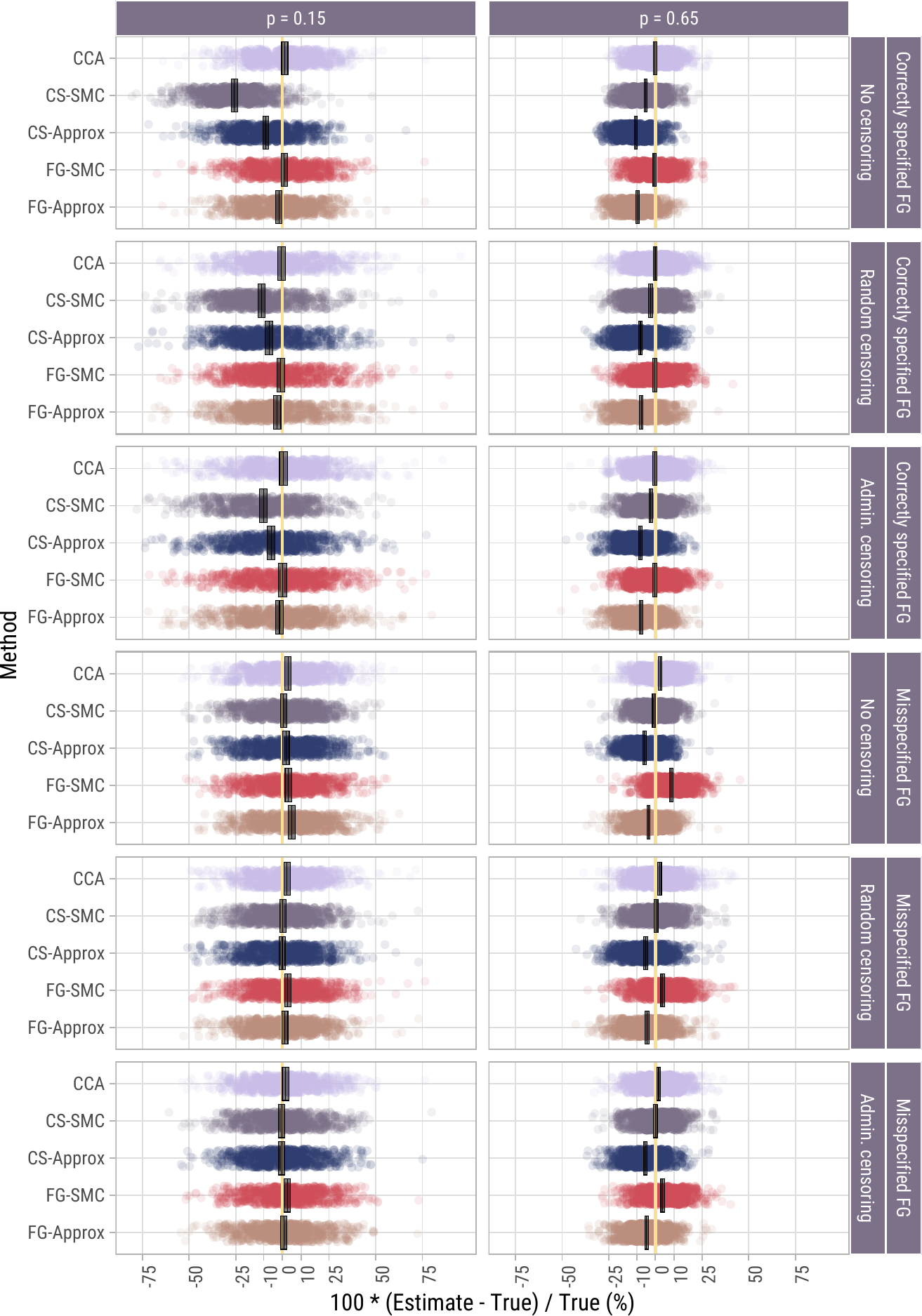}
	\caption{Relative bias (\%) in estimating $\beta_1$. For the correctly specified Fine--Gray (FG) scenarios, $\beta_1 = 0.75$. In the misspecified FG scenarios, the value of the ``least-false'' $\tilde{\beta}_1$ (time-averaged log subdistribution hazard ratio) depended on both $p$ and the presence/absence of censoring. For $p = 0.15$, $\tilde{\beta}_1 \approx 0.76$ without censoring, and $\tilde{\beta}_1 \approx 0.93$ with censoring. For $p = 0.65$, $\tilde{\beta}_1 \approx 0.75$ both with and without censoring. The rectangles drawn for each method and scenario represent the estimated relative bias (vertical line in the middle of a rectangle), and its 95\% Monte Carlo confidence interval (lower and upper limits are represented by the left and right rectangle borders, respectively). The confidence interval is constructed using the standard normal approximation.} 
	\label{fig:bias_X}
\end{figure}

We focus on the results for $\beta_1$, together with its time-averaged analogue $\tilde{\beta}_1$ in the scenarios with time-dependent subdistribution hazard ratios. Results concerning bias are summarised in Figure \ref{fig:bias_X} for all 12 scenarios, and presented on the relative scale (Monte Carlo standard errors were below the desired 0.01 for both bias and relative bias, across all methods and scenarios). 

When the Fine--Gray model for cause 1 was correctly specified, the proposed FG-SMC approach was unbiased regardless of censoring type or (baseline) proportion of cause 1 failures. In contrast, imputing compatibly with the (incorrect) assumption of proportional cause-specific hazards showed strong biases, particularly when $p = 0.15$ in the absence of censoring (25\% biased). In the presence of censoring however, this bias dropped to approximately 10\%. The CS-Approx method showed consistent downward biases regardless of $p$ and censoring type, while the FG-Approx method was only biased when $p = 0.65$. The latter finding is consistent with previous research in the simple survival setting; namely that the approximately compatible MI approach is expected to work well when cumulative incidence is low\cite{whiteImputingMissingCovariate2009}. When the DGM generated event times under proportional cause-specific hazards, the magnitude of any biases present were in general smaller (e.g.~closer to the 5\% mark for approximate MI approaches when $p = 0.65$). For the FG-SMC approach, bias was most noticeable when $p = 0.65$, and in the absence of censoring. CS-SMC was unbiased throughout these misspecified Fine--Gray scenarios.

\begin{figure}[ht]
	\includegraphics[width=0.85\textwidth]{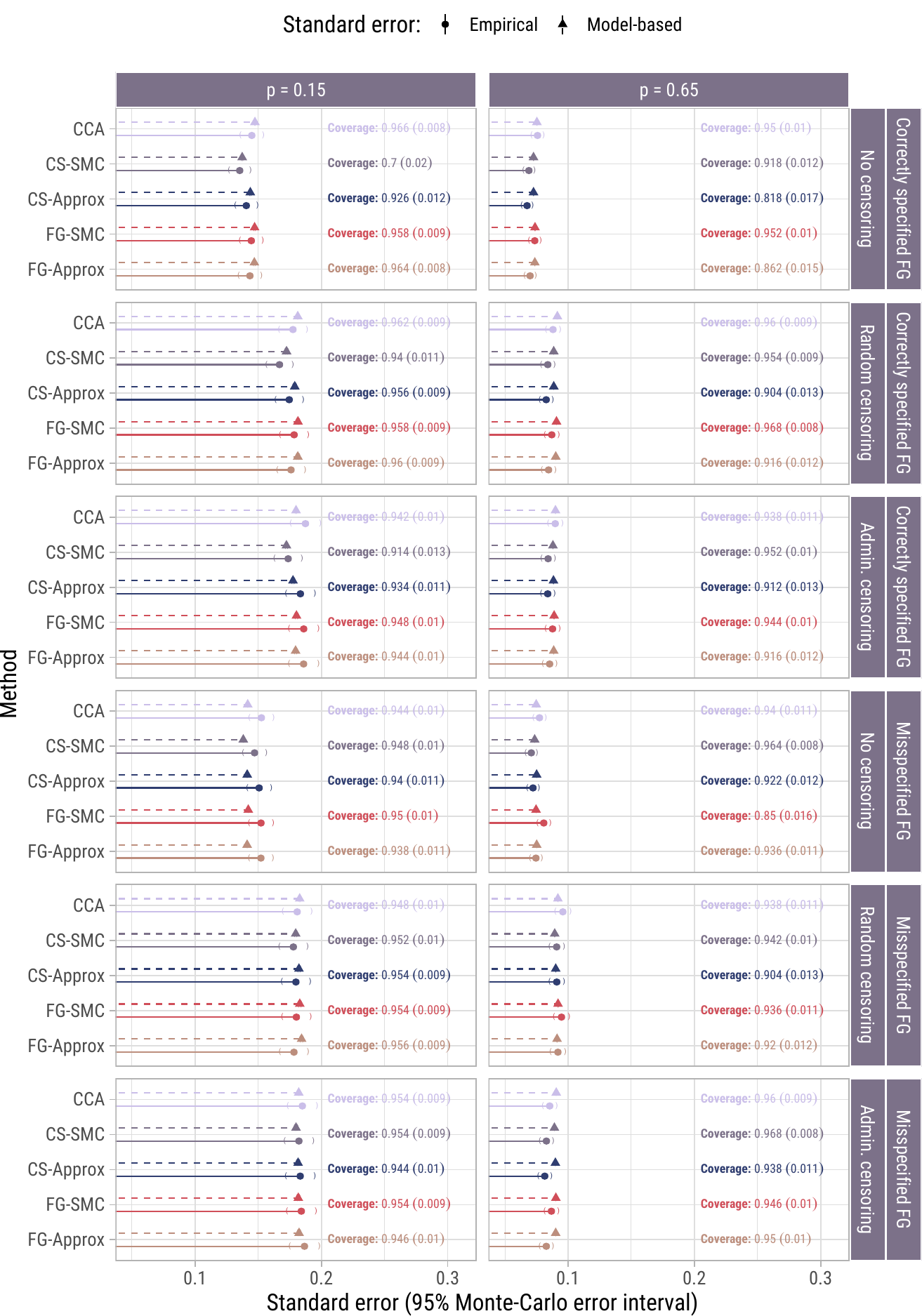}
	\caption{Summary of empirical and model-based standard errrors, together with coverage probabilities for $\beta_1$ (or $\tilde{\beta}_1$ in scenarios with non-proportional subdistribution hazards). Monte Carlo standard errors (MCSEs) are numerically given in brackets for the coverage probabilities, while the MCSEs for model-based/empirical standard errors are shown by 95\% confidence intervals (using standard normal approximation) around a given point. The MCSEs for model-based standard errors are smaller than the graphical width of the point itself, and thus cannot be seen.} 
	\label{fig:perf_X}
\end{figure}

Figure \ref{fig:perf_X} summarises empirical and model-based standard errors, together with coverage probabilities for $\beta_1$ and $\tilde{\beta}_1$. The model-based standard errors were on average close to their empirical counterparts. CS-SMC appears to have a slight variance advantage over competing approaches, mainly when $p = 0.15$. Interestingly, there was no gain in efficiency when the censoring times were known compared to when they needed to be imputed. This is in line with simulation results in both Fine and Gray\cite{fineProportionalHazardsModel1999} and Ruan and Gray\cite{ruanAnalysesCumulativeIncidence2008}, that compared the censoring complete variance estimator (of subdistribution log hazard ratios) to estimators based on the weighted score function and KM imputation method, respectively. The FG-SMC approach showed good coverage (near the nominal 95\% mark) when the Fine--Gray model was correctly specified, although there was slight over-coverage when imputation of censoring times was required. Using the non-parametric bootstrap when estimating $P(C > t)$, which was not investigated in the simulation study, is unlikely to correct for this over-coverage. Under-coverage showed by competing approaches was primarily due to biased estimates.

\subsubsection{Individual-specific cumulative incidences}

\begin{figure}[ht]
	\centering
	\includegraphics[width=0.92\textwidth]{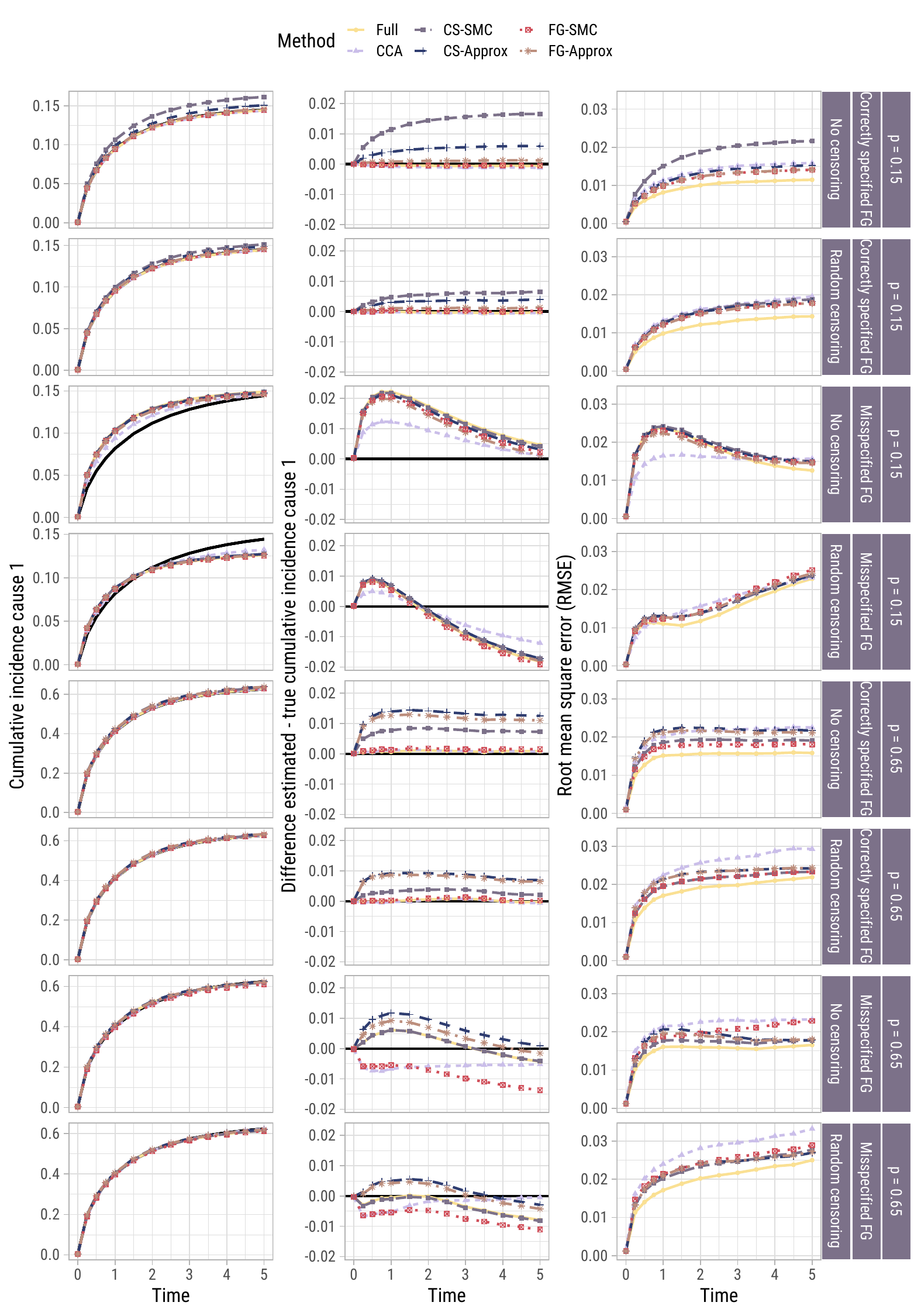}
	\caption{Per scenario (row) for a baseline individual $\{X, Z\} = \{0,0\}$: true (black line) versus estimated cumulative incidence over time, averaged across the 500 replications per scenario (left column); difference between estimated and true (middle column); root mean square error (RMSE) of these estimates (right column). Results for scenarios with administrative censoring are omitted since they were indistinguishable from those with random censoring.} 
	\label{fig:preds_baseline}
\end{figure}

\begin{figure}[ht]
	\centering
	\includegraphics[width=0.92\textwidth]{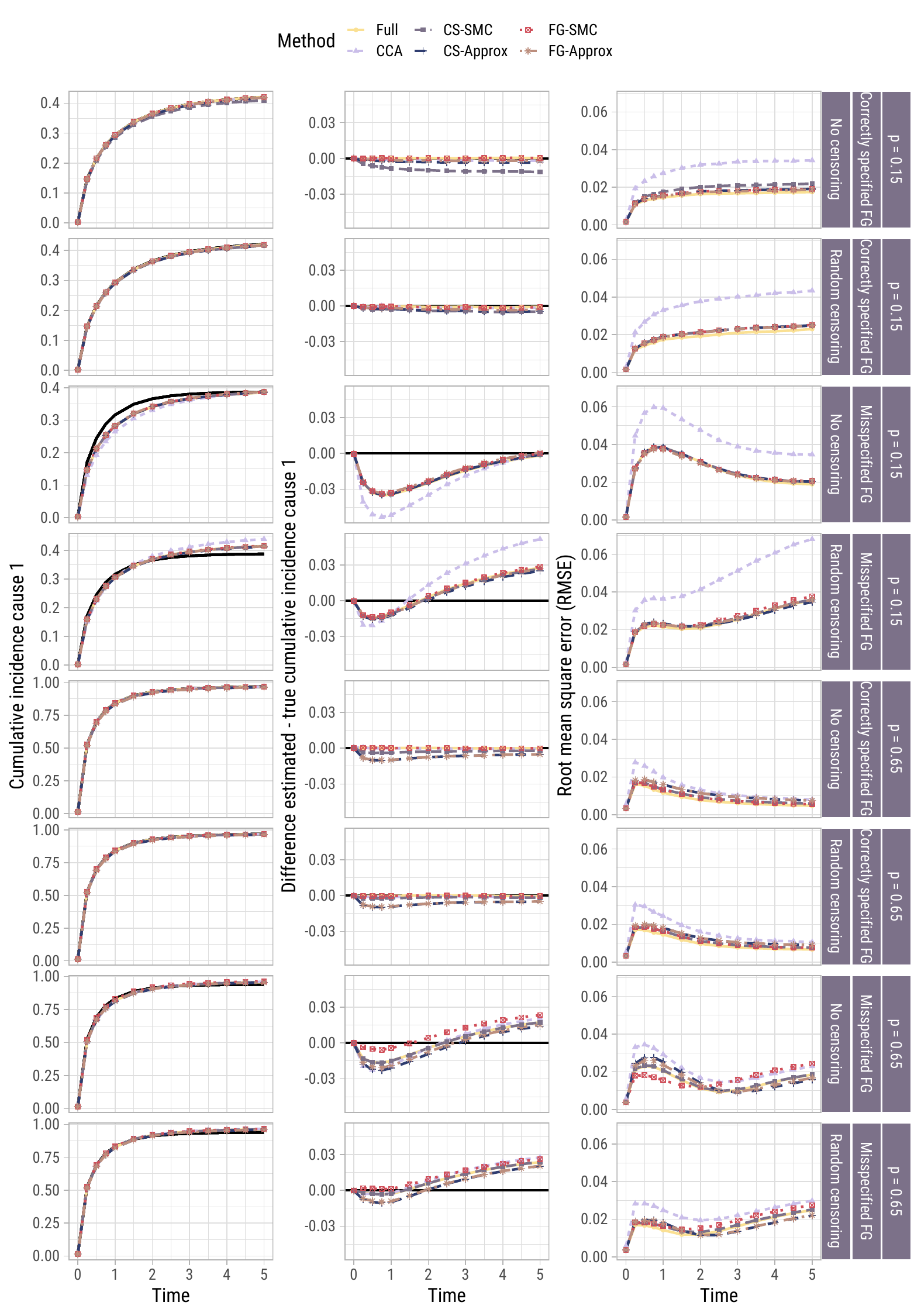}
	\caption{Per scenario (row) for individual $\{X,Z\} = \{1,1\}$: estimated versus true (black line) cumulative incidence over time, averaged across the 500 replications per scenario (left column); difference between estimated and true (middle column); root mean square error (RMSE) of these estimates (right column). Results for scenarios with administrative censoring are omitted since they were indistinguishable from those with random censoring.} 
	\label{fig:preds_X1Z1}
\end{figure}

Figure \ref{fig:preds_baseline} shows the true and average estimated baseline cumulative incidence function $F_0(t)$, the average difference between true and estimated $F_0(t)$, and the RMSE of the estimates. Figure \ref{fig:preds_X1Z1} presents the same information instead for a patient with $\{X, Z\} = \{1, 1\}$. Scenarios where the censoring times are known are omitted from the Figure, as results were indistinguishable from scenarios where the censoring times needed to be imputed. 

The cost of imputing compatibly with the wrong model (using CS-SMC when the Fine--Gray model was correctly specified, or FG-SMC when the DGM was based on cause-specific proportional hazards) when estimating $F_0(t)$ was only noticeable for the CS-SMC approach in the absence of censoring when $p = 0.15$, in terms of both absolute bias and RMSE. On the whole, the approximately compatible MI approaches performed comparably in terms of RMSE to the SMC approaches. In scenarios where the Fine--Gray model was misspecified, the effect of substantive model misspecification (post-imputation) was clear to see in terms of estimating $F_0(t)$ (over- and underestimation at different points in time). When $\{X, Z\} = \{1, 1\}$, all imputation approaches outperformed CCA in terms of RMSE when estimating $F_1(t \given X = 1, Z =1)$, though to a lesser extent when $p = 0.65$. This can presumably be attributed to the efficiency gain in estimating $\beta_2$.

\section{Applied data example}
\label{sec:polverelli}

We illustrate the methods assessed in the simulations study on a dataset of 3982 adult patients with primary and secondary myelofibrosis undergoing alloHCT between 2009 and 2019, and registered with the EBMT \citep{polverelliImpactComorbiditiesBody2024a}. Myelofibrosis is a rare and chronic myeloproliferative neoplasm characterised by bone marrow fibrosis and extramedullary hematopoiesis, for which an alloHCT is the only curative treatment \citep{krogerIndicationManagementAllogeneic2024}. In the original study, the primary objective was to evaluate the association between comorbidities at time of alloHCT and (cause-specific) death without prior relapse of the underlying disease, the so-called non-relapse mortality. In the present illustration, we instead assume that interest lies in developing a prognostic model for time to disease relapse in the first 60 months following an alloHCT. To this end, we developed a Fine--Gray model for relapse, with death prior to relapse as sole competing risk.

A set of 18 baseline predictors were chosen on the basis of substantive clinical knowledge, many of which had a considerable proportion of missing data (see supplementary material S2.1). These predictors included the 13 variables used in the multivariable models from the original study, and 5 additional variables that were either known to be predictive of disease relapse (use of T-cell depletion; presence of cytogenetic abnormalities), or provided relevant auxiliary information regarding the missing values (year of transplantation; time between diagnosis and transplantation; and whether diagnosis was primary or secondary myelofibrosis). Note that since this is a model for (complementary log-log transformed) cumulative incidence of relapse, we want to make sure to include predictors known to be associated with the cause-specific hazards of \textit{both} relapse and non-relapse mortality.

Since around 45\% of patients were either event-free or censored within the first 60 months (see supplementary material S2.2, non-parametric curves), potential censoring times for those experiencing non-relapse mortality were first multiply imputed using the \{kmi\} package in strata defined by (completely observed) year of transplantation, yielding 100 datasets with ``complete'' subdistribution time $V$ but with partially observed covariate information. In each of these datasets, covariates were imputed once using each of the four imputation methods used in the simulation study, after 20 cycles across the covariates. The choice of 100 imputed datasets was motivated using von Hippel's quadratic rule (i.e.~number of imputed datasets needed should increase approximately quadratically with increasing fraction of missing information), based on an initial set of 30 imputed datasets \citep{vonhippelHowManyImputations2020}. Essentially, we sought to control the MCSEs of the standard errors of the estimated subdistribution log hazard ratios. Default imputation methods were used depending on the type of covariate: binary covariates using logistic regression, ordered categorical using proportional odds regression and nominal categorical using multinomial logistic regression. For continuous covariates, the default in \{mice\} is predictive mean matching, while linear regression is used for $f(X_j \given X_{-j}, Z;\psi)$ in \{smcfcs\}. The imputation model for a given partially observed variable therefore contained as predictors all remaining fully and partially observed variables from the substantive model, together with the outcome. Each imputation approach differs mainly in how they incorporate the outcome in the imputation model: either by sampling directly from an assumed substantive model compatible distribution (FG-SMC and CS-SMC), or by including event indicator(s) and marginal cumulative hazard(s) explicitly as additional predictors (FG-Approx and CS-Approx).

\begin{figure}[ht]
	\centering
	\includegraphics[width=\textwidth]{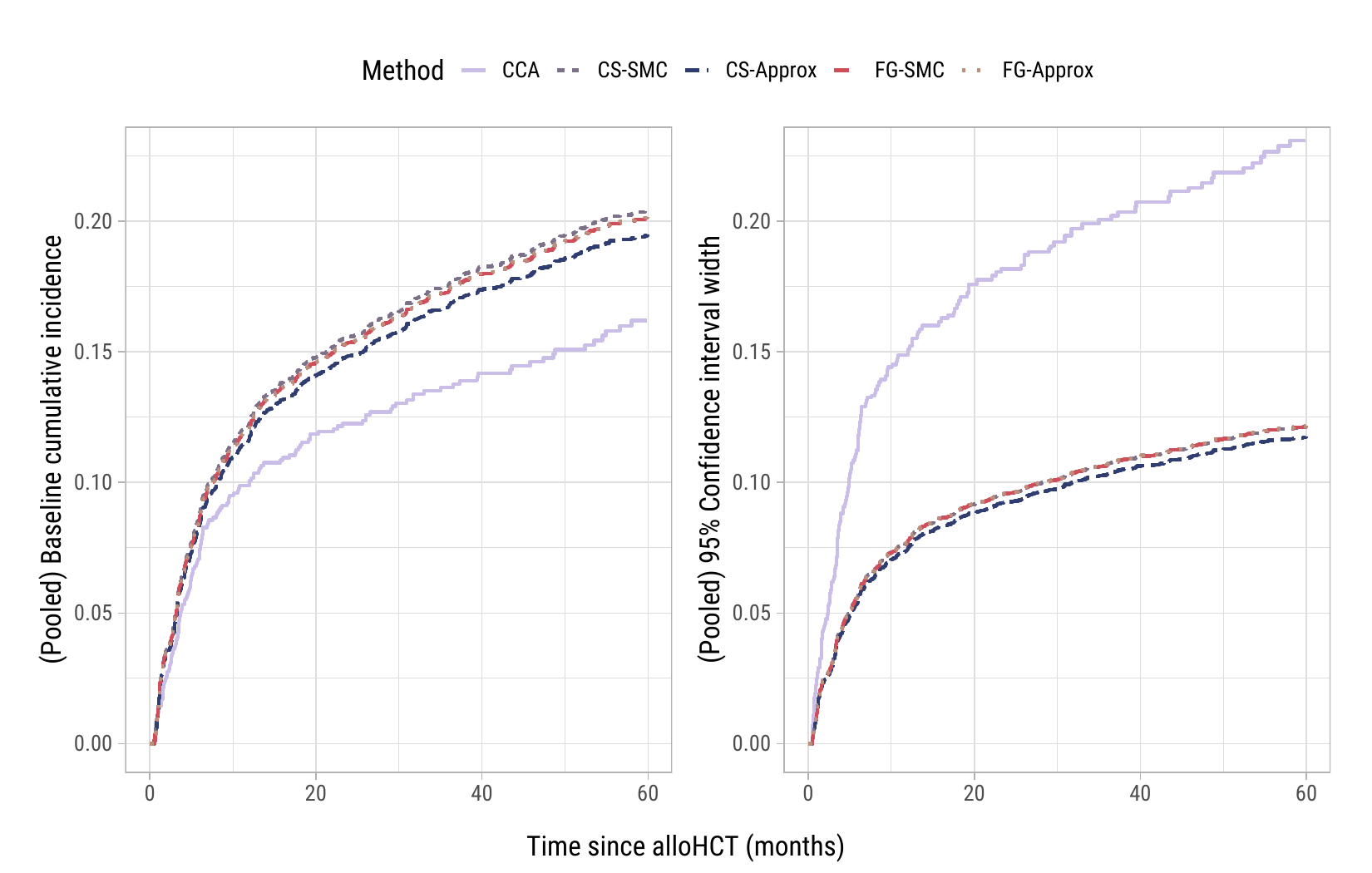}
	\caption{Pooled baseline cumulative incidence functions for relapse in the applied data example (left panel), and width of corresponding confidence intervals (right panel). These are the estimates for a patient aged 60, transplanted in 2019 immediately after diagnosis, with 10g/dL hemoglobin, 15 x10$^9$/L white blood cells, no peripheral blood blasts, and with reference levels for all categorical predictors (see supplementary Table 1).} 
	\label{fig:applied_base_cuminc}
\end{figure}

Figure \ref{fig:applied_base_cuminc} shows for all methods the estimated baseline cumulative incidence function, and the width of the corresponding confidence interval at each timepoint. As was the case in the simulation study, cumulative incidences are estimated in each imputed dataset, and pooled after complementary log-log transformation. The estimation procedure used for the standard errors of the cumulative incidences is described by Ozenne et al.\cite{RJ-2017-062}. The estimates using both FG-SMC and FG-Approx are virtually overlapping, which is consistent with the simulation study results when $p = 0.15$. Both CS-SMC and CS-Approx also yielded cumulative incidences that were close to those obtained by the subdistribution hazard based imputation approaches, which is in line with the results of the simulation study under random right censoring. The most stark differences were between CCA (which only uses 20\% of patients) and the imputation approaches: the cumulative incidence of relapse at 60 months was almost 5\% lower than the nearest MI-based curve, with confidence intervals that were over twice as wide. For completeness, in supplementary material S2.3 we report the pooled subdistribution log hazard ratios, in addition to the pooled coefficients of cause-specific Cox models for relapse and non-relapse mortality (each containing the same predictors as the Fine--Gray model for relapse). The pooled coefficients of the Fine--Gray models were extremely similar between imputation approaches, and all differed considerably from the (much more variable) CCA. There were some noticeable differences between subdistribution hazard based and cause-specific hazard based imputation approaches when estimating the cause-specific Cox model for non-relapse mortality (see e.g.~pooled coefficients for weight loss prior to transplantation, hemoglobin or high risk comorbidity score). Furthermore, the pooled subdistribution log hazard ratios were generally small in magnitude (none exceeding 0.5), a setting in which both SMC and approximately compatible approaches are expected to perform similarly. 

The differences observed between point estimates obtained using the imputation based approaches and CCA are in large part explainable by the gulf in efficiency between the two approaches. Nevertheless, there are indications that the estimates obtained using imputation methods would be less biased than their CCA counterparts in this example. An exploratory logistic model showed that the observed time to competing event and competing event indicator were both predictive of the probability of being a complete-case, after adjusting for other known important predictors of missingness such as year of transplantation (many variables recorded more often later on in time as their clinical relevance became clearer). Upon closer inspection, it appears that the probability of being a complete-case is significantly lower only for those censored earlier on in time. This seemingly unlikely association between future outcome and baseline complete-case indicator (outcome-dependent MAR, under which CCA is biased) is likely confounded by transplant centre. That is, shorter follow-up times and missing values in covariates may both be symptomatic of a given centre's overall quality of data collection. Although ignored in the present analysis for simplicity, there is indeed heterogeneity in data completeness between EBMT affiliated transplant centres across and within different countries. The MI of potential censoring times would allow to model centre effects using standard software, for example by means of stratification or use of a frailty term.

\section{Discussion} \label{sec:discussion}

In this paper, we extended the SMC-FCS approach in order to impute missing covariates compatibly with a Fine--Gray substantive model. For a given competing event, the theory relies on using the subdistribution time $V$ and the corresponding event-specific indicator as outcome variables. In the presence of random right-censoring, $V$ is only partially observed, as the potential censoring times for those failing from competing events are unknown. These can be multiply imputed in a first step, after which covariates can be imputed by conditioning on the ``complete'' outcome variables. The approach is straightforward to implement in R by making use of existing software packages \{kmi\} and \{smcfcs\}. While the imputation of potential censoring times appears underused in the subdistribution hazard modelling literature (relative to weighted approaches), it has inspired other methodological extensions e.g.~enabling the use of deep learning in discrete time after single imputation of potential censoring times\cite{gorgizadehImputationApproachUsing2022a}.

The simulation study compared the performance of the proposed method to competing MI approaches, including imputing compatibly with cause-specific proportional hazards models. The FG-SMC approach performed optimally (in terms of estimating both subdistribution log hazard ratios, and cumulative incidences) when the assumption of proportional subdistribution hazards held, and performed satisfactorily when this assumption did not hold. For cumulative incidence estimation, the choice of substantive model (i.e.~cause-specific Cox vs.~Fine--Gray) at the analysis phase appears to be more important than the procedure used to impute the missing covariates. In terms of RMSE of these predictions, most imputation approaches outperform CCA. The applied data example also demonstrated the possible gain in efficiency when using MI instead of CCA. 

One counterintuitive finding was that the presence of censoring seems to \textit{improve} the performance of the misspecified SMC-FCS procedure (e.g.~use of CS-SMC when underlying DGM assumes proportional subdistribution hazards). An explanation for this phenomenon is that the time-dependent factor relating the cause-specific and subdistribution hazards for cause 1 (the ``reduction factor''\cite{putterRelationCausespecificHazard2020a}) is closer to 1 earlier in time. Therefore (in the example with DGM assuming proportional subdistribution hazards), the violation of proportionality on the cause-specific hazard scale will appear to be less severe in earlier time-periods, thereby improving the performance of the misspecified SMC-FCS approach. This is also in line with earlier findings showing how similar the results of subdistribution and cause-specific hazards models can be in presence of heavy censoring\cite{grambauerProportionalSubdistributionHazards2010a,vanderpasDifferentCompetingRisks2018a}.

An advantage to the proposed approach is that it can be extended in various ways. First, while not yet possible using existing \{smcfcs\} software, the censoring times can depend on partially observed $X$ (rather than just on complete $Z$), meaning $X$ and $V$ will need to be imputed iteratively. The FG-Approx method with censoring depending on $X$ can be easily implemented in \{mice\} using custom imputation methods. Second, the approach can account for time-dependent effects, by making direct use of existing approaches developed in the context of standard Cox models\cite{keoghMultipleImputationCox2018b}. Third, the proposed approach can be extended to accommodate interval censored outcomes, using the methodology described by Delord and G\'{e}nin, which relies on analogous principles: multiply impute interval censored $V$ in order to work with simpler censoring complete data\cite{delordMultipleImputationCompeting2016}.

There are multiple limitations to the present work. The first is that the proposed SMC-FCS approach does not accommodate delayed entry (left truncation). Our current recommendation to impute approximately compatibly with a Fine--Gray model subject to delayed entry and right-censoring is to include $I(D=1)$ and $\hat{\Lambda}_1(T)$ as predictors in the imputation model, in addition to other substantive model covariates. Here, $\hat{\Lambda}_1(t)$ is the estimated cumulative subdistribution hazard based on a marginal model that uses time-dependent weights in order to accommodate both left-truncation and right-censoring \citep{geskusCauseSpecificCumulativeIncidence2011}. Note the proposed imputation model uses $\hat{\Lambda}_1(T)$ and not $\hat{\Lambda}_1(V)$, and therefore some downward bias is to be expected, as explained in Appendix \ref{sec:appendix_subdist}. Second, while FG-SMC does not require an explicit model for the competing risks, it does require the censoring distribution to be specified explicitly (e.g.~non-parametrically using KM, or using a Cox model). Third, the proposed approach is geared towards imputing missing covariates when only one competing event is of interest. More generally, the strategy of estimating a Fine--Gray for each cause in turn is not an approach the current authors endorse, based on both theoretical \citep{beyersmannCompetingRisksMultistate2012,AustinTFP2021} and simulation-based arguments \citep{bonnevilleWhyYouShould2024}. When multiple competing events are of interest, we would instead recommend modelling the cause-specific hazards, or using the semiparametric approach suggested by Mao and Lin for joint inference on the cumulative incidence functions\cite{maoEfficientEstimationSemiparametric2017}.

In conclusion, the proposed approach is most appropriate for imputing missing covariates in the context of prognostic modelling of only one event of interest. Based on the simulation study, imputing compatibly with cause-specific proportional hazards seems to be a good all-round strategy for a ``complete'' competing risks analysis (investigating both the cause-specific hazards and cumulative incidence functions\cite{latoucheCompetingRisksAnalysis2013}), and can at the same time be used for prognostic modelling based on the cause-specific Cox models.

\section*{Software}

All R code (needed to reproduce simulation study, applied data example, and manuscript figures) is available at \url{https://github.com/survival-lumc/FineGrayCovarMI}. In addition to the minimal R code provided in the supplementary materials, a wrapper function for the proposed SMC-FCS Fine--Gray method is available inside the \{smcfcs\} R package. 

\section*{Acknowledgements}

The authors would like to thank the Chronic Malignancies Working Party (CMWP) of the EBMT for allowing the use of the myelofibrosis dataset, as well as the patients and centres involved in the original study. We also thank Linda Koster (EBMT Leiden Study Unit) for continued support with this dataset. This work was supported by Grant BE 4500/4-1 from the German Research Foundation (JB), Grant W222158-2-50 from the Leiden University Fund (EFB), and UK MRC grant MR/T023953/2 (JWB).

\section*{Conflicts of interest}

The authors declare no conflicts of interest.


\appendix

\section[Imputed censoring times, and resulting cumulative subdistribution hazards]{\texorpdfstring{Imputed censoring times, and resulting cumulative subdistribution hazards}} \label{sec:appendix_subdist}

\begin{figure}[ht]
	\centering
	\includegraphics[width=\textwidth]{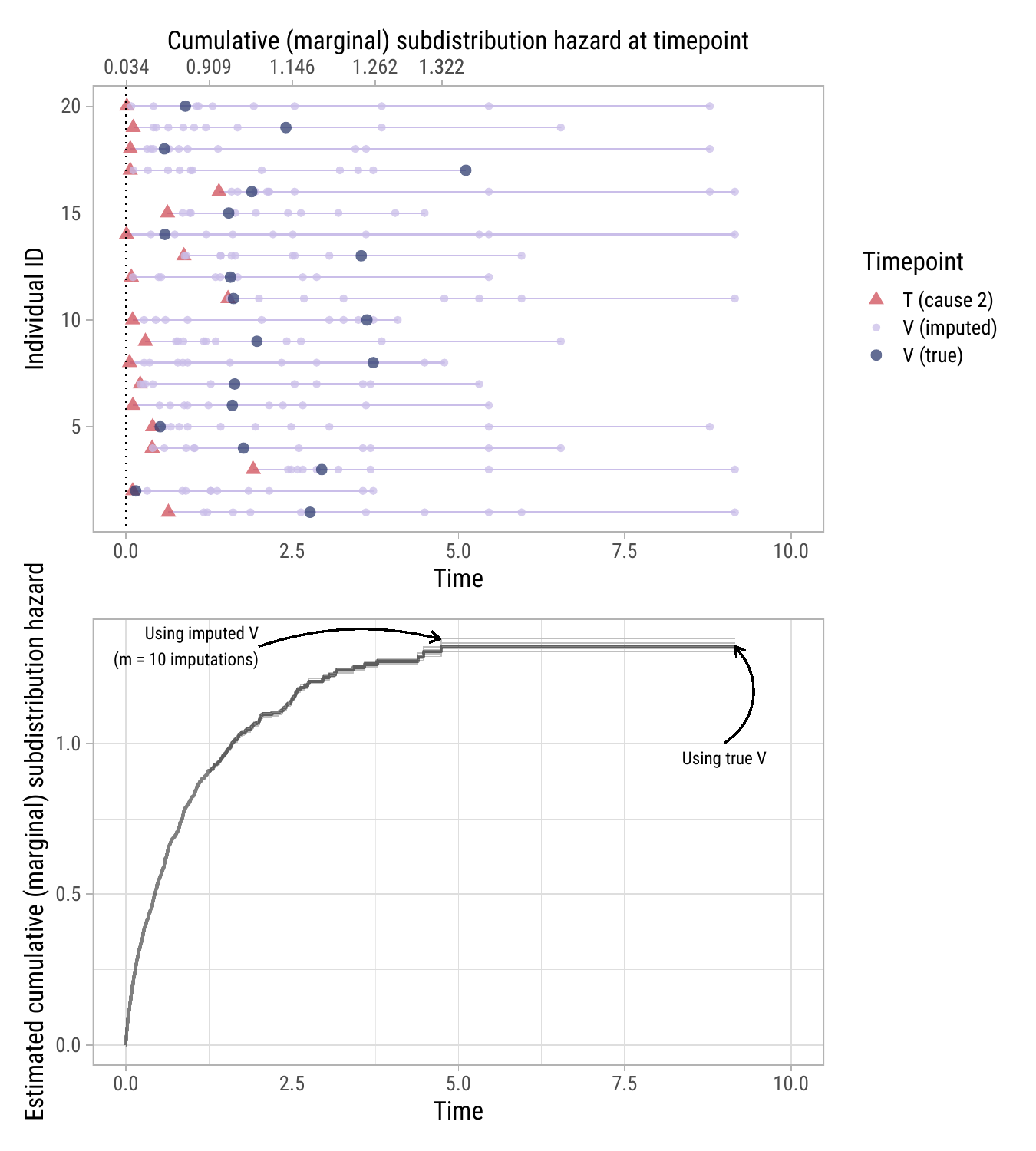}
	\caption{Based on a simulated dataset of $n = 2000$ (correctly specified Fine--Gray, $p = 0.65$, random censoring), we show the imputed ($m = 10$ imputations) potential censoring times for a random selection of 20 individuals failing from cause 2 (upper panel); and the estimated marginal cumulative subdistribution hazard function for cause 1 based on true $V$, and based on imputed $V$ (lower panel).} 
	\label{fig:cens_appendix}
\end{figure}

As described in Section \ref{ssec:imp_cens}, the subdistribution time $V$ is only partially observed in the presence of random right-censoring. Thus, the potential censoring times for those failing from cause 2 should first be multiply imputed, before imputing any missing covariates. This imputation of partially observed $V$ is visualised more closely in Figure \ref{fig:cens_appendix}, using a simulated dataset of 2000 individuals following the parametrisation used in the simulation study scenario with correctly specified Fine--Gray, $p = 0.65$, and random exponential censoring. In this example, the potential censoring times for those failing from cause 2 were imputed $m = 10$ times.

The upper panel shows the imputed potential censoring times for a random selection of 20 individuals failing from cause 2, in addition to their cause 2 failure time and their true eventual censoring time. The lower panel shows the estimated marginal cumulative subdistribution hazard function for $\hat{\Lambda}_1(t)$ resulting from using $I(D = 1)$ together with either the imputed or true $V$ as outcomes in a marginal model. We used $\hat{\Lambda}_1(t)$ estimated using the true $V$ to create the secondary x-axis in the upper panel, which shows the value of this function at a given timepoint. For example, the marginal cumulative subdistribution hazard was 1.146 at timepoint 2.5, and stayed constant at 1.322 after the last cause 1 event in this sample.  

The upper panel in particular gives additional insights regarding the FG-Approx method, where $I(D = 1)$ and $\hat{\Lambda}_1(V)$ are included as predictors in the imputation model. Namely, the secondary x-axis shows the value of $\hat{\Lambda}_1(V)$ used in the imputation model for a missing $X_j$, for given imputed $V$. A first key point is that one should always use $\hat{\Lambda}_1(V)$ in the imputation model, and not $\hat{\Lambda}_1(T)$. Since the observed cause 2 failure time occurs before the eventual censoring time, $\hat{\Lambda}_1(T)$ will always be smaller than the marginal cumulative subdistribution hazard at the eventual censoring time, and so the imputation model will incur some downward bias. A second point is that in settings with fewer event 1 failures (e.g.~$p = 0.15$ scenario in the simulation study), the corresponding secondary x-axis will have a smaller range, since the subdistribution hazard will be lower overall. The imputed potential censoring times will therefore have a relatively smaller influence on the imputed $X_j$.

The lower panel shows that the estimated $\hat{\Lambda}_1(t)$ varies very little between imputed datasets, with differences only being noticeable later on in follow-up as risk sets become smaller and associated cumulative hazard jumps more pronounced. Note also that while $\hat{\Lambda}_1(t)$ based on the true $V$ appears in this dataset to be a kind of ``average'' of the functions based on imputed $V$, this will not be the case in general, especially with smaller sample sizes. The $\hat{\Lambda}_1(t)$ based on the weighted estimator\cite{geskusCauseSpecificCumulativeIncidence2011} will however coincide with the ``average'' of the functions based on imputed $V$, as will using the negative log of one minus the Aalen-Johansen estimate of the marginal cumulative incidence function.

\end{document}


\maketitle
\hypertarget{minimal-code-example}{%
\section{Minimal code example}\label{minimal-code-example}}

This is the minimal \texttt{R} code companion to section 3.4 of main
manuscript. The parameters from the simulation study scenario with
\(p = 0.15\), random censoring, and correctly specified Fine--Gray were
used to generate the example dataset below.

\begin{Shaded}
\begin{Highlighting}[]
\CommentTok{\# Load libraries}
\FunctionTok{library}\NormalTok{(data.table)}
\FunctionTok{library}\NormalTok{(survival)}
\FunctionTok{library}\NormalTok{(kmi)}
\FunctionTok{library}\NormalTok{(mice)}
\FunctionTok{library}\NormalTok{(smcfcs)}

\CommentTok{\# Minimal dataset}
\FunctionTok{head}\NormalTok{(dat, }\AttributeTok{n =} \DecValTok{10}\NormalTok{)}
\end{Highlighting}
\end{Shaded}

\begin{verbatim}
   id     time D    X      Z
1   1 0.491195 0    1  0.126
2   2 0.028680 2 <NA>  1.266
3   3 0.910797 0    0 -1.571
4   4 0.217566 2    1 -0.500
5   5 0.132420 2    0  0.781
6   6 0.800913 2    0 -0.434
7   7 0.041653 2 <NA> -0.844
8   8 0.036202 1 <NA>  1.564
9   9 0.046798 0    0 -1.653
10 10 0.997413 0 <NA> -1.196
\end{verbatim}

\begin{Shaded}
\begin{Highlighting}[]
\FunctionTok{sapply}\NormalTok{(dat, class)}
\end{Highlighting}
\end{Shaded}

\begin{verbatim}
       id      time         D         X         Z 
"integer" "numeric"  "factor"  "factor" "numeric" 
\end{verbatim}

\begin{Shaded}
\begin{Highlighting}[]
\FunctionTok{nrow}\NormalTok{(dat)}
\end{Highlighting}
\end{Shaded}

\begin{verbatim}
[1] 2000
\end{verbatim}

\begin{enumerate}
\def\labelenumi{\arabic{enumi}.}
\tightlist
\item
  Add columns \(\hat{H}_1(T)\) and \(\hat{H}_2(T)\) to the original
  data, which are the marginal cause-specific cumulative hazards for
  each competing risk evaluated at an individual's event or censoring
  time (obtained using the Nelson--Aalen estimator).
\end{enumerate}

\begin{Shaded}
\begin{Highlighting}[]
\CommentTok{\# Add cause{-}specific event indicators + cumulative hazards}
\NormalTok{dat}\SpecialCharTok{$}\NormalTok{D1 }\OtherTok{\textless{}{-}} \FunctionTok{as.numeric}\NormalTok{(dat}\SpecialCharTok{$}\NormalTok{D }\SpecialCharTok{==} \DecValTok{1}\NormalTok{)}
\NormalTok{dat}\SpecialCharTok{$}\NormalTok{D2 }\OtherTok{\textless{}{-}} \FunctionTok{as.numeric}\NormalTok{(dat}\SpecialCharTok{$}\NormalTok{D }\SpecialCharTok{==} \DecValTok{2}\NormalTok{)}
\NormalTok{dat}\SpecialCharTok{$}\NormalTok{H1 }\OtherTok{\textless{}{-}} \FunctionTok{nelsonaalen}\NormalTok{(}\AttributeTok{data =}\NormalTok{ dat, }\AttributeTok{timevar =} \StringTok{"time"}\NormalTok{, }\AttributeTok{statusvar =} \StringTok{"D1"}\NormalTok{)}
\NormalTok{dat}\SpecialCharTok{$}\NormalTok{H2 }\OtherTok{\textless{}{-}} \FunctionTok{nelsonaalen}\NormalTok{(}\AttributeTok{data =}\NormalTok{ dat, }\AttributeTok{timevar =} \StringTok{"time"}\NormalTok{, }\AttributeTok{statusvar =} \StringTok{"D2"}\NormalTok{)}
\end{Highlighting}
\end{Shaded}

\begin{enumerate}
\def\labelenumi{\arabic{enumi}.}
\setcounter{enumi}{1}
\tightlist
\item
  Multiply impute the potential censoring for those failing from cause 2
  using \{kmi\}, yielding \(m\) censoring complete datasets (i.e.~with
  ``complete'' \(V\)). Any completely observed covariates that are known
  to affect the probability of being censored should be included as
  predictors in the model for the censoring process. \{kmi\} imputes
  based on stratified Kaplan--Meier when \(Z\) are categorical, and
  based on a Cox model when at least one of \(Z\) are continuous.
\end{enumerate}

\begin{Shaded}
\begin{Highlighting}[]
\CommentTok{\# 5 imputed datasets}
\NormalTok{M }\OtherTok{\textless{}{-}} \DecValTok{5}

\CommentTok{\# Multiply impute the censoring times}
\NormalTok{cens\_imps }\OtherTok{\textless{}{-}} \FunctionTok{kmi}\NormalTok{(}
  \AttributeTok{formula =} \FunctionTok{Surv}\NormalTok{(time, D }\SpecialCharTok{!=} \DecValTok{0}\NormalTok{) }\SpecialCharTok{\textasciitilde{}} \DecValTok{1}\NormalTok{, }\CommentTok{\# Additional predictors added here}
  \AttributeTok{data =}\NormalTok{ dat,}
  \AttributeTok{etype =}\NormalTok{ D,}
  \AttributeTok{failcode =} \DecValTok{1}\NormalTok{, }\CommentTok{\# Specify event of interest}
  \AttributeTok{nimp =}\NormalTok{ M}
\NormalTok{)}
\end{Highlighting}
\end{Shaded}

\begin{enumerate}
\def\labelenumi{\arabic{enumi}.}
\setcounter{enumi}{2}
\tightlist
\item
  In each censoring complete dataset, add an additional column
  \(\hat{\Lambda}_1(V)\). This takes the value of the marginal
  cumulative subdistribution hazard for cause 1 at an individual's
  observed or imputed subdistribution time, obtained with the
  Nelson--Aalen estimator based on \(I(D = 1)\) and imputed \(V\).
\end{enumerate}

\begin{Shaded}
\begin{Highlighting}[]
\CommentTok{\# Preparation for covariate imputation: }
\CommentTok{\# Create list of censoring complete datasets (with imputed V)}
\NormalTok{list\_to\_impute }\OtherTok{\textless{}{-}} \FunctionTok{lapply}\NormalTok{(cens\_imps}\SpecialCharTok{$}\NormalTok{imputed.data, }\ControlFlowTok{function}\NormalTok{(imp\_dat) \{}
  
  \CommentTok{\# Adjust new ordering from kmi (cause 2 individuals appended at bottom)}
\NormalTok{  dat\_to\_impute }\OtherTok{\textless{}{-}} \FunctionTok{cbind}\NormalTok{(cens\_imps}\SpecialCharTok{$}\NormalTok{original.data, imp\_dat)}
  
  \CommentTok{\# Compute/add Lambda\_1(V) in each imputed dataset}
\NormalTok{  dat\_to\_impute}\SpecialCharTok{$}\NormalTok{Lambda1 }\OtherTok{\textless{}{-}} \FunctionTok{nelsonaalen}\NormalTok{(}
    \AttributeTok{data =}\NormalTok{ dat\_to\_impute, }
    \AttributeTok{timevar =} \StringTok{"newtimes"}\NormalTok{, }\CommentTok{\# kmi naming for V}
    \AttributeTok{statusvar =} \StringTok{"D1"} \CommentTok{\# I(D=1)}
\NormalTok{  )}
  \FunctionTok{return}\NormalTok{(dat\_to\_impute)}
\NormalTok{\})}

\CommentTok{\# newevent is equal to I(D=1)}
\FunctionTok{head}\NormalTok{(list\_to\_impute[[}\DecValTok{1}\NormalTok{]])}
\end{Highlighting}
\end{Shaded}

\begin{verbatim}
   id     time D    X      Z D1 D2         H1         H2 newtimes newevent
1   1 0.491195 0    1  0.126  0  0 0.16736459 0.55436927 0.491195        0
3   3 0.910797 0    0 -1.571  0  0 0.25761243 0.83833716 0.910797        0
8   8 0.036202 1 <NA>  1.564  1  0 0.02028935 0.09603222 0.036202        1
9   9 0.046798 0    0 -1.653  0  0 0.02606228 0.10990397 0.046798        0
10 10 0.997413 0 <NA> -1.196  0  0 0.27549886 0.87116320 0.997413        0
12 12 0.056015 0 <NA>  0.058  0  0 0.02903112 0.12350351 0.056015        0
      Lambda1
1  0.12385222
3  0.16659793
8  0.01932257
9  0.02452308
10 0.17340532
12 0.02715245
\end{verbatim}

\begin{enumerate}
\def\labelenumi{\arabic{enumi}.}
\setcounter{enumi}{3}
\tightlist
\item
  In each censoring complete dataset (each with different \(V\) and
  \(\hat{\Lambda}_1(V)\), but same \(\hat{H}_1(T)\) and
  \(\hat{H}_2(T)\)), create a single imputed dataset using the desired
  covariate imputation method(s).
\end{enumerate}

\begin{Shaded}
\begin{Highlighting}[]
\CommentTok{\# Prepare predictor matrices for MICE using first censoring complete dataset}
\NormalTok{predmat\_cs\_approx }\OtherTok{\textless{}{-}}\NormalTok{ predmat\_fg\_approx }\OtherTok{\textless{}{-}}\NormalTok{ mice}\SpecialCharTok{::}\FunctionTok{make.predictorMatrix}\NormalTok{(}
  \AttributeTok{data =}\NormalTok{ list\_to\_impute[[}\DecValTok{1}\NormalTok{]]}
\NormalTok{)}
\NormalTok{predmat\_cs\_approx[] }\OtherTok{\textless{}{-}}\NormalTok{ predmat\_fg\_approx[] }\OtherTok{\textless{}{-}} \DecValTok{0}

\CommentTok{\# Explicitly specify predictors to include in the imputation model}
\NormalTok{predmat\_cs\_approx[}\StringTok{"X"}\NormalTok{, }\FunctionTok{c}\NormalTok{(}\StringTok{"Z"}\NormalTok{, }\StringTok{"D1"}\NormalTok{, }\StringTok{"D2"}\NormalTok{, }\StringTok{"H1"}\NormalTok{, }\StringTok{"H2"}\NormalTok{)] }\OtherTok{\textless{}{-}} \DecValTok{1}
\NormalTok{predmat\_fg\_approx[}\StringTok{"X"}\NormalTok{, }\FunctionTok{c}\NormalTok{(}\StringTok{"Z"}\NormalTok{, }\StringTok{"D1"}\NormalTok{, }\StringTok{"Lambda1"}\NormalTok{)] }\OtherTok{\textless{}{-}} \DecValTok{1}
\NormalTok{predmat\_fg\_approx}
\end{Highlighting}
\end{Shaded}

\begin{verbatim}
         id time D X Z D1 D2 H1 H2 newtimes newevent Lambda1
id        0    0 0 0 0  0  0  0  0        0        0       0
time      0    0 0 0 0  0  0  0  0        0        0       0
D         0    0 0 0 0  0  0  0  0        0        0       0
X         0    0 0 0 1  1  0  0  0        0        0       1
Z         0    0 0 0 0  0  0  0  0        0        0       0
D1        0    0 0 0 0  0  0  0  0        0        0       0
D2        0    0 0 0 0  0  0  0  0        0        0       0
H1        0    0 0 0 0  0  0  0  0        0        0       0
H2        0    0 0 0 0  0  0  0  0        0        0       0
newtimes  0    0 0 0 0  0  0  0  0        0        0       0
newevent  0    0 0 0 0  0  0  0  0        0        0       0
Lambda1   0    0 0 0 0  0  0  0  0        0        0       0
\end{verbatim}

\begin{Shaded}
\begin{Highlighting}[]
\CommentTok{\# Prepare the methods:}
\CommentTok{\# {-} Approx methods: model type for X | Z, outcome}
\NormalTok{methods\_approx }\OtherTok{\textless{}{-}}\NormalTok{ mice}\SpecialCharTok{::}\FunctionTok{make.method}\NormalTok{(}\AttributeTok{data =}\NormalTok{ list\_to\_impute[[}\DecValTok{1}\NormalTok{]])}

\CommentTok{\# {-} SMC methods: proposal model for X | Z (need to use \{smcfcs\} naming)}
\NormalTok{methods\_smcfcs }\OtherTok{\textless{}{-}}\NormalTok{ mice}\SpecialCharTok{::}\FunctionTok{make.method}\NormalTok{(}
  \AttributeTok{data =}\NormalTok{ list\_to\_impute[[}\DecValTok{1}\NormalTok{]],}
  \AttributeTok{defaultMethod =} \FunctionTok{c}\NormalTok{(}\StringTok{"norm"}\NormalTok{, }\StringTok{"logreg"}\NormalTok{, }\StringTok{"mlogit"}\NormalTok{, }\StringTok{"podds"}\NormalTok{)}
\NormalTok{)}
\NormalTok{methods\_smcfcs}
\end{Highlighting}
\end{Shaded}

\begin{verbatim}
      id     time        D        X        Z       D1       D2       H1 
      ""       ""       "" "logreg"       ""       ""       ""       "" 
      H2 newtimes newevent  Lambda1 
      ""       ""       ""       "" 
\end{verbatim}

\begin{Shaded}
\begin{Highlighting}[]
\CommentTok{\# Impute X in each censoring complete dataset}
\CommentTok{\# (parallelise this loop for speed improvements on larger data)}
\NormalTok{list\_imps }\OtherTok{\textless{}{-}} \FunctionTok{lapply}\NormalTok{(list\_to\_impute, }\ControlFlowTok{function}\NormalTok{(imp\_dat) \{}

\NormalTok{  m }\OtherTok{\textless{}{-}} \DecValTok{1}
\NormalTok{  iters }\OtherTok{\textless{}{-}} \DecValTok{10}
  
\NormalTok{  imps\_cs\_approx }\OtherTok{\textless{}{-}} \FunctionTok{mice}\NormalTok{(}
    \AttributeTok{data =}\NormalTok{ imp\_dat,}
    \AttributeTok{m =}\NormalTok{ m,}
    \AttributeTok{maxit =}\NormalTok{ iters,}
    \AttributeTok{method =}\NormalTok{ methods\_approx,}
    \AttributeTok{predictorMatrix =}\NormalTok{ predmat\_cs\_approx}
\NormalTok{  )}

\NormalTok{  imps\_fg\_approx }\OtherTok{\textless{}{-}} \FunctionTok{mice}\NormalTok{(}
    \AttributeTok{data =}\NormalTok{ imp\_dat,}
    \AttributeTok{m =}\NormalTok{ m,}
    \AttributeTok{maxit =}\NormalTok{ iters,}
    \AttributeTok{method =}\NormalTok{ methods\_approx,}
    \AttributeTok{predictorMatrix =}\NormalTok{ predmat\_fg\_approx}
\NormalTok{  )}

\NormalTok{  imps\_cs\_smc }\OtherTok{\textless{}{-}} \FunctionTok{smcfcs}\NormalTok{(}
    \AttributeTok{originaldata =}\NormalTok{ imp\_dat,}
    \AttributeTok{smtype =} \StringTok{"compet"}\NormalTok{,}
    \AttributeTok{smformula =} \FunctionTok{list}\NormalTok{(}
      \StringTok{"Surv(time, D == 1) \textasciitilde{} X + Z"}\NormalTok{,}
      \StringTok{"Surv(time, D == 2) \textasciitilde{} X + Z"}
\NormalTok{    ),}
    \AttributeTok{method =}\NormalTok{ methods\_smcfcs,}
    \AttributeTok{m =}\NormalTok{ m,}
    \AttributeTok{numit =}\NormalTok{ iters}
\NormalTok{  )}

\NormalTok{  imps\_fg\_smc }\OtherTok{\textless{}{-}} \FunctionTok{smcfcs}\NormalTok{(}
    \AttributeTok{originaldata =}\NormalTok{ imp\_dat,}
    \AttributeTok{smtype =} \StringTok{"coxph"}\NormalTok{,}
    \AttributeTok{smformula =} \StringTok{"Surv(newtimes, D1) \textasciitilde{} X + Z"}\NormalTok{,}
    \AttributeTok{method =}\NormalTok{ methods\_smcfcs,}
    \AttributeTok{m =}\NormalTok{ m,}
    \AttributeTok{numit =}\NormalTok{ iters}
\NormalTok{  )}

  \CommentTok{\# Bring all the imputed datasets together}
\NormalTok{  imps }\OtherTok{\textless{}{-}} \FunctionTok{rbind.data.frame}\NormalTok{(}
    \FunctionTok{cbind}\NormalTok{(}\AttributeTok{method =} \StringTok{"CCA"}\NormalTok{, imp\_dat),}
    \FunctionTok{cbind}\NormalTok{(}\AttributeTok{method =} \StringTok{"cs\_smc"}\NormalTok{, imps\_cs\_smc}\SpecialCharTok{$}\NormalTok{impDatasets[[}\DecValTok{1}\NormalTok{]]),}
    \FunctionTok{cbind}\NormalTok{(}\AttributeTok{method =} \StringTok{"cs\_approx"}\NormalTok{, }\FunctionTok{complete}\NormalTok{(imps\_cs\_approx, }\AttributeTok{action =}\NormalTok{ 1L)),}
    \FunctionTok{cbind}\NormalTok{(}\AttributeTok{method =} \StringTok{"fg\_smc"}\NormalTok{, imps\_fg\_smc}\SpecialCharTok{$}\NormalTok{impDatasets[[}\DecValTok{1}\NormalTok{]]),}
    \FunctionTok{cbind}\NormalTok{(}\AttributeTok{method =} \StringTok{"fg\_approx"}\NormalTok{, }\FunctionTok{complete}\NormalTok{(imps\_cs\_approx, }\AttributeTok{action =}\NormalTok{ 1L))}
\NormalTok{  )}
  \FunctionTok{return}\NormalTok{(imps)}
\NormalTok{\})}
\end{Highlighting}
\end{Shaded}

\begin{enumerate}
\def\labelenumi{\arabic{enumi}.}
\setcounter{enumi}{4}
\tightlist
\item
  Fit the Fine--Gray substantive model in each imputed dataset (using
  standard Cox software with \(I(D = 1)\) and imputed \(V\) as outcome
  variables), and pool the estimates using Rubin's rules.
\end{enumerate}

\begin{Shaded}
\begin{Highlighting}[]
\CommentTok{\# Bind everything together}
\NormalTok{dat\_imps }\OtherTok{\textless{}{-}} \FunctionTok{rbindlist}\NormalTok{(list\_imps, }\AttributeTok{idcol =} \StringTok{".imp"}\NormalTok{)}
\NormalTok{dat\_imps}
\end{Highlighting}
\end{Shaded}

\begin{verbatim}
       .imp    method   id     time D    X      Z D1 D2         H1         H2
    1:    1       CCA    1 0.491195 0    1  0.126  0  0 0.16736459 0.55436927
    2:    1       CCA    3 0.910797 0    0 -1.571  0  0 0.25761243 0.83833716
    3:    1       CCA    8 0.036202 1 <NA>  1.564  1  0 0.02028935 0.09603222
    4:    1       CCA    9 0.046798 0    0 -1.653  0  0 0.02606228 0.10990397
    5:    1       CCA   10 0.997413 0 <NA> -1.196  0  0 0.27549886 0.87116320
   ---                                                                       
49996:    5 fg_approx 1992 0.319702 2    0 -2.670  0  1 0.12370372 0.43826433
49997:    5 fg_approx 1993 0.229071 2    0 -0.243  0  1 0.09740419 0.35023923
49998:    5 fg_approx 1994 1.836303 2    1 -0.366  0  1 0.47538639 1.23075745
49999:    5 fg_approx 1997 0.702380 2    0  0.283  0  1 0.21877205 0.71087168
50000:    5 fg_approx 1999 0.023554 2    1  1.377  0  1 0.01356742 0.06584427
       newtimes newevent    Lambda1
    1: 0.491195        0 0.12385222
    2: 0.910797        0 0.16659793
    3: 0.036202        1 0.01932257
    4: 0.046798        0 0.02452308
    5: 0.997413        0 0.17340532
   ---                             
49996: 0.957205        0 0.17116627
49997: 0.453168        0 0.12098105
49998: 2.841599        0 0.25988878
49999: 1.170590        0 0.19454317
50000: 2.997529        0 0.26284736
\end{verbatim}

\begin{Shaded}
\begin{Highlighting}[]
\CommentTok{\# To use the usual workflow: subset one of the methods first}
\NormalTok{imps\_fg\_smc }\OtherTok{\textless{}{-}}\NormalTok{ dat\_imps[dat\_imps}\SpecialCharTok{$}\NormalTok{method }\SpecialCharTok{==} \StringTok{"fg\_smc"}\NormalTok{, ]}

\CommentTok{\# Fit model in each imputed dataset}
\NormalTok{mods\_fg\_smc }\OtherTok{\textless{}{-}} \FunctionTok{lapply}\NormalTok{(}
  \AttributeTok{X =} \FunctionTok{seq\_len}\NormalTok{(M), }
  \AttributeTok{FUN =} \ControlFlowTok{function}\NormalTok{(m) \{}
\NormalTok{    imp\_m }\OtherTok{\textless{}{-}}\NormalTok{ imps\_fg\_smc[imps\_fg\_smc}\SpecialCharTok{$}\NormalTok{.imp }\SpecialCharTok{==}\NormalTok{ m, ]}
    \FunctionTok{coxph}\NormalTok{(}\FunctionTok{Surv}\NormalTok{(newtimes, D1) }\SpecialCharTok{\textasciitilde{}}\NormalTok{ X }\SpecialCharTok{+}\NormalTok{ Z, }\AttributeTok{data =}\NormalTok{ imp\_m)}
\NormalTok{  \} }
\NormalTok{)}

\CommentTok{\# Pool results}
\FunctionTok{summary}\NormalTok{(}\FunctionTok{pool}\NormalTok{(mods\_fg\_smc))}
\end{Highlighting}
\end{Shaded}

\begin{verbatim}
  term  estimate  std.error statistic         df      p.value
1   X1 0.7768682 0.21722362  3.576352   9.883541 5.136286e-03
2    Z 0.4920664 0.06519244  7.547906 105.385333 1.659276e-11
\end{verbatim}

\begin{Shaded}
\begin{Highlighting}[]
\CommentTok{\# Alternative: }
\CommentTok{\# Use (nested) \{data.table\} workflow to pool all methods simultaneously!}
\NormalTok{dat\_mods }\OtherTok{\textless{}{-}}\NormalTok{ dat\_imps[, .(}
  \AttributeTok{mod =} \FunctionTok{list}\NormalTok{(}\FunctionTok{coxph}\NormalTok{(}\FunctionTok{Surv}\NormalTok{(newtimes, D1) }\SpecialCharTok{\textasciitilde{}}\NormalTok{ X }\SpecialCharTok{+}\NormalTok{ Z, }\AttributeTok{data =}\NormalTok{ .SD))}
\NormalTok{), by }\OtherTok{=} \FunctionTok{c}\NormalTok{(}\StringTok{"method"}\NormalTok{, }\StringTok{".imp"}\NormalTok{)]}
\NormalTok{dat\_mods}
\end{Highlighting}
\end{Shaded}

\begin{verbatim}
       method .imp         mod
 1:       CCA    1 <coxph[22]>
 2:    cs_smc    1 <coxph[21]>
 3: cs_approx    1 <coxph[21]>
 4:    fg_smc    1 <coxph[21]>
 5: fg_approx    1 <coxph[21]>
 6:       CCA    2 <coxph[22]>
 7:    cs_smc    2 <coxph[21]>
 8: cs_approx    2 <coxph[21]>
 9:    fg_smc    2 <coxph[21]>
10: fg_approx    2 <coxph[21]>
11:       CCA    3 <coxph[22]>
12:    cs_smc    3 <coxph[21]>
13: cs_approx    3 <coxph[21]>
14:    fg_smc    3 <coxph[21]>
15: fg_approx    3 <coxph[21]>
16:       CCA    4 <coxph[22]>
17:    cs_smc    4 <coxph[21]>
18: cs_approx    4 <coxph[21]>
19:    fg_smc    4 <coxph[21]>
20: fg_approx    4 <coxph[21]>
21:       CCA    5 <coxph[22]>
22:    cs_smc    5 <coxph[21]>
23: cs_approx    5 <coxph[21]>
24:    fg_smc    5 <coxph[21]>
25: fg_approx    5 <coxph[21]>
       method .imp         mod
\end{verbatim}

\begin{Shaded}
\begin{Highlighting}[]
\NormalTok{dat\_mods[, }\FunctionTok{summary}\NormalTok{(}\FunctionTok{pool}\NormalTok{(}\FunctionTok{as.list}\NormalTok{(mod))), by }\OtherTok{=} \StringTok{"method"}\NormalTok{]}
\end{Highlighting}
\end{Shaded}

\begin{verbatim}
       method term  estimate  std.error statistic         df      p.value
 1:       CCA   X1 0.7781281 0.17916465  4.343089 152.067624 2.554742e-05
 2:       CCA    Z 0.4003856 0.10186017  3.930737 145.744472 1.304356e-04
 3:    cs_smc   X1 0.6980657 0.18538543  3.765483  14.973349 1.875994e-03
 4:    cs_smc    Z 0.5079436 0.06538007  7.769090  93.531830 9.965454e-12
 5: cs_approx   X1 0.6092265 0.19461615  3.130400  12.205414 8.525728e-03
 6: cs_approx    Z 0.5225790 0.06779656  7.708046  58.618467 1.775328e-10
 7:    fg_smc   X1 0.7768682 0.21722362  3.576352   9.883541 5.136286e-03
 8:    fg_smc    Z 0.4920664 0.06519244  7.547906 105.385333 1.659276e-11
 9: fg_approx   X1 0.6092265 0.19461615  3.130400  12.205414 8.525728e-03
10: fg_approx    Z 0.5225790 0.06779656  7.708046  58.618467 1.775328e-10
\end{verbatim}

\newpage

\hypertarget{applied-data-example}{%
\section{Applied data example}\label{applied-data-example}}

\hypertarget{data-dictionary}{%
\subsection{Data dictionary}\label{data-dictionary}}

\begin{longtable}[t]{lc}
\caption{Data dictionary. CMV: cytomegalovirus; HLA: human leukocyte antigen; HCT-CI: Hematopoietic stem cell transplantation-comorbidity index; MF: myelofibrosis.}\\
\toprule
Characteristic & N = 3,982\\
\midrule
Patient age (years) & 58 (52, 64)\\
Patient/donor CMV match & \\
\hspace{1em}Patient negative/Donor negative & 1,142 (30\%)\\
\hspace{1em}Other & 2,715 (70\%)\\
\hspace{1em}(Missing) & 125\\
Donor type & \\
\hspace{1em}HLA identical sibling & 1,183 (30\%)\\
\hspace{1em}Other & 2,795 (70\%)\\
\hspace{1em}(Missing) & 4\\
Hemoglobin (g/dL) & 9.10 (8.10, 10.40)\\
\hspace{1em}(Missing) & 1,873\\
HCT-CI risk category & \\
\hspace{1em}Low risk ($0$) & 1,674 (54\%)\\
\hspace{1em}Intermediate risk ($1-2$) & 743 (24\%)\\
\hspace{1em}High risk ($\geq 3$) & 674 (22\%)\\
\hspace{1em}(Missing) & 891\\
Interval diagnosis-transplantation (years) & 3 (1, 9)\\
Karnosfky performance score & \\
\hspace{1em}$\geq 90$ & 2,475 (66\%)\\
\hspace{1em}$80$ & 986 (26\%)\\
\hspace{1em}$\leq 70$ & 267 (7.2\%)\\
\hspace{1em}(Missing) & 254\\
Patient sex & \\
\hspace{1em}Female & 1,484 (37\%)\\
\hspace{1em}Male & 2,498 (63\%)\\
Peripheral blood (PB) blasts (\%) & 1.0 (0.0, 3.0)\\
\hspace{1em}(Missing) & 2,323\\
Conditioning & \\
\hspace{1em}Standard & 1,373 (35\%)\\
\hspace{1em}Reduced & 2,553 (65\%)\\
\hspace{1em}(Missing) & 56\\
Ruxolitinib given & \\
\hspace{1em}No & 1,832 (66\%)\\
\hspace{1em}Yes & 931 (34\%)\\
\hspace{1em}(Missing) & 1,219\\
Disease subclassification & \\
\hspace{1em}Primary MF & 2,912 (73\%)\\
\hspace{1em}Secondary MF & 1,070 (27\%)\\
Night sweats & \\
\hspace{1em}No & 1,256 (70\%)\\
\hspace{1em}Yes & 529 (30\%)\\
\hspace{1em}(Missing) & 2,197\\
T-cell depletion (in- or ev-vivo) & \\
\hspace{1em}No & 1,012 (26\%)\\
\hspace{1em}Yes & 2,905 (74\%)\\
\hspace{1em}(Missing) & 65\\
Cytogenetics & \\
\hspace{1em}Normal & 1,318 (59\%)\\
\hspace{1em}Abnormal & 910 (41\%)\\
\hspace{1em}(Missing) & 1,754\\
White blood cell count (WBC, x$10^9$/L) & 7 (4, 14)\\
\hspace{1em}(Missing) & 1,884\\
>10\% Weight loss prior to transplantation & \\
\hspace{1em}No & 1,329 (73\%)\\
\hspace{1em}Yes & 492 (27\%)\\
\hspace{1em}(Missing) & 2,161\\
Year of transplantation & 2,015.0 (2,012.0, 2,018.0)\\
\bottomrule
\multicolumn{2}{l}{\rule{0pt}{1em}\textsuperscript{1} Median (IQR); n (\%)}\\
\end{longtable}

\hypertarget{non-parametric-cumulative-incidence-curves}{%
\subsection{Non-parametric cumulative incidence
curves}\label{non-parametric-cumulative-incidence-curves}}

\begin{figure}[H]

{\centering \includegraphics{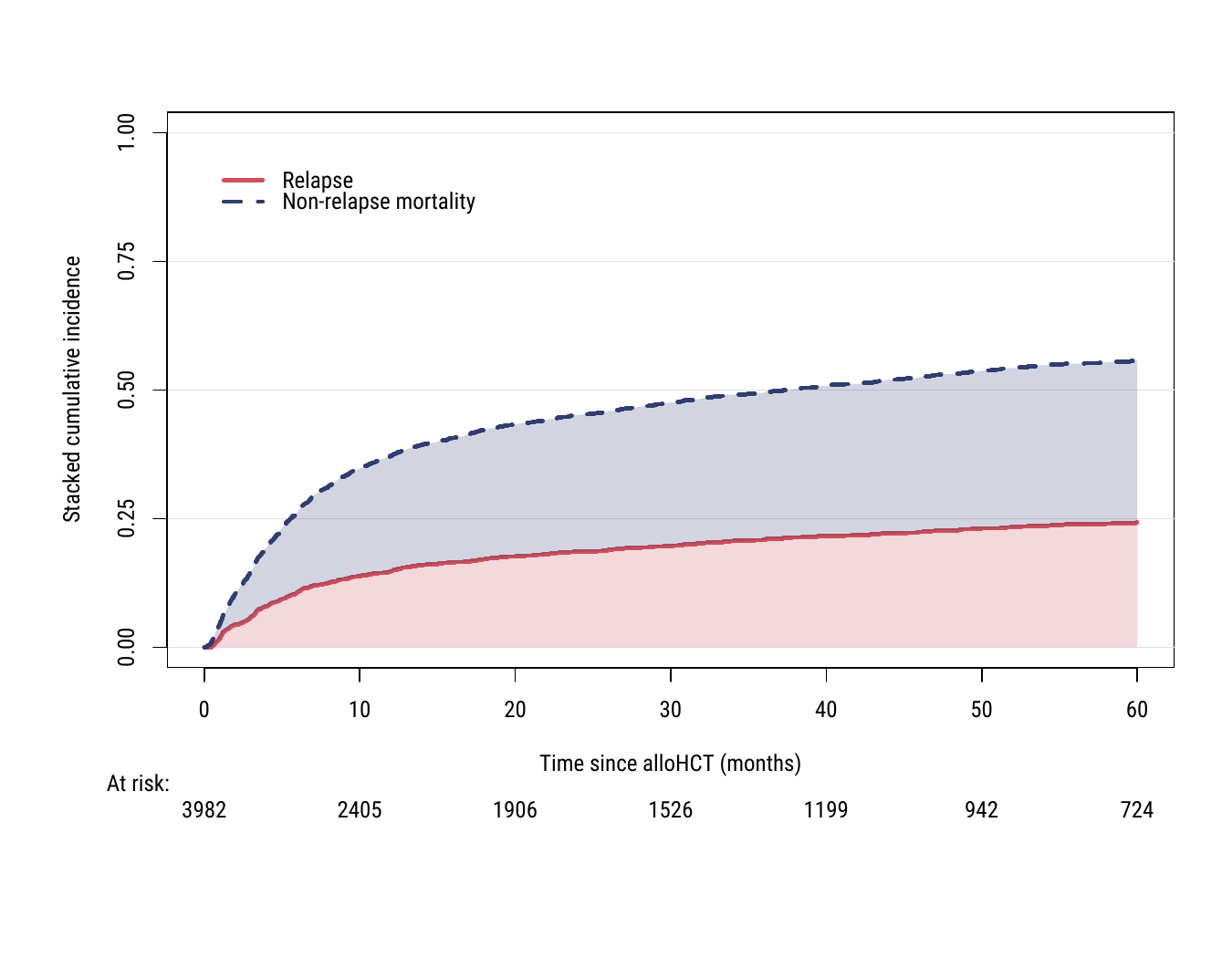}

}

\caption{Stacked non-parametric cumulative incidence curves for
competing relapse and non-relapse mortality, in dataset of 3982 primary
and secondary myelofibrosis patients.}

\end{figure}

\hypertarget{pooled-regression-coefficeints}{%
\subsection{Pooled regression
coefficeints}\label{pooled-regression-coefficeints}}

\begingroup\fontsize{9}{11}\selectfont

\begin{longtable}[t]{lrrr}
\caption{Pooled log hazard ratios [log HR, 95\% confidence interval] for Fine--Gray model for relapse, cause-specific Cox model relapse, and cause-specific Cox model for non-relapse mortality (NRM).}\\
\toprule
Term + method & Relapse subdist.~log HR & Relapse cause-spec.~log HR & NRM cause-spec.~log HR\\
\midrule
\endfirsthead
\caption[]{\textit{(continued)}}\\
\toprule
Term + method & Relapse subdist.~log HR & Relapse cause-spec.~log HR & NRM cause-spec.~log HR\\
\midrule
\endhead
\midrule
\multicolumn{4}{r@{}}{\textit{(continued \ldots)}}\\
\endfoot
\bottomrule
\endlastfoot
\addlinespace[0.3em]
\multicolumn{4}{l}{\textbf{Conditioning: reduced}}\\
\hspace{1em}CCA & 0.02 [-0.33, 0.36] & 0.01 [-0.33, 0.35] & 0 [-0.29, 0.28]\\
\hspace{1em}CS-SMC & 0.13 [-0.02, 0.28] & 0.1 [-0.05, 0.25] & -0.05 [-0.18, 0.07]\\
\hspace{1em}CS-Approx & 0.13 [-0.02, 0.28] & 0.1 [-0.05, 0.25] & -0.05 [-0.18, 0.07]\\
\hspace{1em}FG-SMC & 0.13 [-0.02, 0.28] & 0.1 [-0.05, 0.25] & -0.06 [-0.18, 0.07]\\
\hspace{1em}FG-Approx & 0.13 [-0.03, 0.28] & 0.1 [-0.06, 0.25] & -0.05 [-0.18, 0.07]\\
\addlinespace[0.3em]
\multicolumn{4}{l}{\textbf{CMV match: other}}\\
\hspace{1em}CCA & 0.04 [-0.31, 0.4] & 0.05 [-0.3, 0.41] & 0.09 [-0.19, 0.37]\\
\hspace{1em}CS-SMC & -0.1 [-0.26, 0.05] & -0.05 [-0.2, 0.11] & 0.22 [0.08, 0.36]\\
\hspace{1em}CS-Approx & -0.1 [-0.26, 0.05] & -0.05 [-0.2, 0.11] & 0.22 [0.08, 0.36]\\
\hspace{1em}FG-SMC & -0.1 [-0.26, 0.05] & -0.04 [-0.2, 0.11] & 0.22 [0.08, 0.36]\\
\hspace{1em}FG-Approx & -0.11 [-0.26, 0.05] & -0.05 [-0.2, 0.11] & 0.22 [0.08, 0.35]\\
\addlinespace[0.3em]
\multicolumn{4}{l}{\textbf{Cytogenetics: abnormal}}\\
\hspace{1em}CCA & 0.36 [0.04, 0.68] & 0.37 [0.05, 0.68] & -0.08 [-0.35, 0.19]\\
\hspace{1em}CS-SMC & 0.35 [0.15, 0.54] & 0.35 [0.16, 0.54] & -0.07 [-0.23, 0.1]\\
\hspace{1em}CS-Approx & 0.36 [0.17, 0.55] & 0.35 [0.16, 0.54] & -0.08 [-0.25, 0.08]\\
\hspace{1em}FG-SMC & 0.36 [0.17, 0.55] & 0.36 [0.17, 0.54] & -0.06 [-0.21, 0.08]\\
\hspace{1em}FG-Approx & 0.34 [0.17, 0.52] & 0.34 [0.17, 0.51] & -0.07 [-0.22, 0.08]\\
\addlinespace[0.3em]
\multicolumn{4}{l}{\textbf{Donor relation: other}}\\
\hspace{1em}CCA & 0.12 [-0.28, 0.52] & 0.2 [-0.2, 0.6] & 0.53 [0.18, 0.88]\\
\hspace{1em}CS-SMC & -0.26 [-0.41, -0.1] & -0.19 [-0.34, -0.03] & 0.35 [0.21, 0.5]\\
\hspace{1em}CS-Approx & -0.25 [-0.41, -0.1] & -0.18 [-0.34, -0.02] & 0.36 [0.21, 0.5]\\
\hspace{1em}FG-SMC & -0.26 [-0.41, -0.1] & -0.19 [-0.34, -0.03] & 0.35 [0.2, 0.49]\\
\hspace{1em}FG-Approx & -0.26 [-0.41, -0.1] & -0.19 [-0.34, -0.03] & 0.35 [0.2, 0.49]\\
\addlinespace[0.3em]
\multicolumn{4}{l}{\textbf{Hemoglobin (per $5$ g/dL)}}\\
\hspace{1em}CCA & -0.38 [-0.85, 0.09] & -0.39 [-0.85, 0.08] & -0.12 [-0.49, 0.25]\\
\hspace{1em}CS-SMC & -0.24 [-0.51, 0.03] & -0.3 [-0.58, -0.03] & -0.19 [-0.42, 0.04]\\
\hspace{1em}CS-Approx & -0.25 [-0.53, 0.02] & -0.32 [-0.59, -0.06] & -0.19 [-0.41, 0.02]\\
\hspace{1em}FG-SMC & -0.25 [-0.51, 0.02] & -0.29 [-0.56, -0.02] & -0.08 [-0.28, 0.11]\\
\hspace{1em}FG-Approx & -0.23 [-0.5, 0.04] & -0.27 [-0.54, 0] & -0.09 [-0.29, 0.11]\\
\addlinespace[0.3em]
\multicolumn{4}{l}{\textbf{HCT-CI ($1-2$)}}\\
\hspace{1em}CCA & -0.15 [-0.53, 0.22] & -0.04 [-0.42, 0.33] & 0.38 [0.08, 0.69]\\
\hspace{1em}CS-SMC & -0.22 [-0.42, -0.01] & -0.17 [-0.37, 0.03] & 0.15 [-0.02, 0.31]\\
\hspace{1em}CS-Approx & -0.19 [-0.38, 0.01] & -0.14 [-0.34, 0.06] & 0.15 [-0.01, 0.31]\\
\hspace{1em}FG-SMC & -0.22 [-0.42, -0.01] & -0.18 [-0.38, 0.02] & 0.12 [-0.04, 0.28]\\
\hspace{1em}FG-Approx & -0.19 [-0.38, 0.01] & -0.15 [-0.35, 0.04] & 0.11 [-0.05, 0.27]\\
\addlinespace[0.3em]
\multicolumn{4}{l}{\textbf{HCT-CI ($\geq 3$)}}\\
\hspace{1em}CCA & -0.27 [-0.7, 0.16] & -0.19 [-0.62, 0.23] & 0.4 [0.07, 0.73]\\
\hspace{1em}CS-SMC & -0.07 [-0.28, 0.14] & -0.01 [-0.21, 0.2] & 0.27 [0.1, 0.44]\\
\hspace{1em}CS-Approx & -0.08 [-0.28, 0.13] & -0.02 [-0.22, 0.18] & 0.26 [0.1, 0.43]\\
\hspace{1em}FG-SMC & -0.06 [-0.27, 0.14] & -0.02 [-0.22, 0.19] & 0.21 [0.05, 0.37]\\
\hspace{1em}FG-Approx & -0.08 [-0.28, 0.11] & -0.04 [-0.23, 0.16] & 0.21 [0.05, 0.38]\\
\addlinespace[0.3em]
\multicolumn{4}{l}{\textbf{Interval diagnosis to alloHCT (decades)}}\\
\hspace{1em}CCA & 0.01 [-0.24, 0.26] & 0 [-0.25, 0.26] & -0.03 [-0.25, 0.19]\\
\hspace{1em}CS-SMC & -0.02 [-0.14, 0.09] & -0.02 [-0.14, 0.1] & 0.05 [-0.05, 0.15]\\
\hspace{1em}CS-Approx & -0.03 [-0.14, 0.09] & -0.02 [-0.14, 0.1] & 0.05 [-0.05, 0.15]\\
\hspace{1em}FG-SMC & -0.02 [-0.14, 0.09] & -0.02 [-0.13, 0.1] & 0.05 [-0.05, 0.15]\\
\hspace{1em}FG-Approx & -0.02 [-0.14, 0.09] & -0.02 [-0.14, 0.1] & 0.05 [-0.05, 0.15]\\
\addlinespace[0.3em]
\multicolumn{4}{l}{\textbf{Karnofsky ($80$)}}\\
\hspace{1em}CCA & -0.09 [-0.48, 0.31] & -0.08 [-0.48, 0.31] & 0.04 [-0.27, 0.34]\\
\hspace{1em}CS-SMC & 0.07 [-0.1, 0.24] & 0.12 [-0.05, 0.28] & 0.17 [0.03, 0.31]\\
\hspace{1em}CS-Approx & 0.06 [-0.1, 0.23] & 0.1 [-0.06, 0.27] & 0.15 [0.01, 0.29]\\
\hspace{1em}FG-SMC & 0.07 [-0.09, 0.24] & 0.12 [-0.05, 0.29] & 0.17 [0.03, 0.31]\\
\hspace{1em}FG-Approx & 0.07 [-0.1, 0.24] & 0.12 [-0.06, 0.29] & 0.17 [0.03, 0.31]\\
\addlinespace[0.3em]
\multicolumn{4}{l}{\textbf{Karnofsky ($\leq 70$)}}\\
\hspace{1em}CCA & 0.63 [0.15, 1.11] & 0.79 [0.3, 1.28] & 0.33 [-0.13, 0.79]\\
\hspace{1em}CS-SMC & 0.44 [0.19, 0.69] & 0.55 [0.3, 0.81] & 0.31 [0.08, 0.53]\\
\hspace{1em}CS-Approx & 0.42 [0.17, 0.67] & 0.51 [0.26, 0.76] & 0.26 [0.04, 0.49]\\
\hspace{1em}FG-SMC & 0.44 [0.19, 0.7] & 0.55 [0.29, 0.81] & 0.32 [0.09, 0.54]\\
\hspace{1em}FG-Approx & 0.43 [0.17, 0.68] & 0.53 [0.28, 0.78] & 0.31 [0.08, 0.53]\\
\addlinespace[0.3em]
\multicolumn{4}{l}{\textbf{Disease subclassification: secondary MF}}\\
\hspace{1em}CCA & -0.05 [-0.45, 0.35] & -0.02 [-0.42, 0.38] & 0.07 [-0.27, 0.41]\\
\hspace{1em}CS-SMC & 0.01 [-0.17, 0.19] & 0.01 [-0.17, 0.19] & 0 [-0.16, 0.15]\\
\hspace{1em}CS-Approx & 0 [-0.18, 0.18] & 0 [-0.18, 0.19] & 0 [-0.16, 0.15]\\
\hspace{1em}FG-SMC & 0 [-0.18, 0.18] & 0 [-0.18, 0.18] & -0.01 [-0.16, 0.15]\\
\hspace{1em}FG-Approx & 0 [-0.18, 0.18] & 0 [-0.18, 0.18] & -0.01 [-0.16, 0.15]\\
\addlinespace[0.3em]
\multicolumn{4}{l}{\textbf{Night sweats: yes}}\\
\hspace{1em}CCA & -0.33 [-0.7, 0.04] & -0.4 [-0.77, -0.02] & -0.02 [-0.32, 0.27]\\
\hspace{1em}CS-SMC & -0.18 [-0.41, 0.05] & -0.2 [-0.44, 0.03] & -0.02 [-0.23, 0.19]\\
\hspace{1em}CS-Approx & -0.12 [-0.36, 0.13] & -0.14 [-0.38, 0.1] & 0.03 [-0.19, 0.24]\\
\hspace{1em}FG-SMC & -0.17 [-0.4, 0.07] & -0.18 [-0.41, 0.05] & 0.01 [-0.16, 0.19]\\
\hspace{1em}FG-Approx & -0.16 [-0.4, 0.07] & -0.18 [-0.42, 0.05] & 0 [-0.17, 0.18]\\
\addlinespace[0.3em]
\multicolumn{4}{l}{\textbf{Patient age (decades)}}\\
\hspace{1em}CCA & 0.1 [-0.09, 0.28] & 0.13 [-0.06, 0.32] & 0.13 [-0.02, 0.28]\\
\hspace{1em}CS-SMC & -0.03 [-0.12, 0.05] & 0.01 [-0.08, 0.09] & 0.21 [0.14, 0.29]\\
\hspace{1em}CS-Approx & -0.03 [-0.12, 0.05] & 0.01 [-0.08, 0.09] & 0.21 [0.14, 0.29]\\
\hspace{1em}FG-SMC & -0.04 [-0.12, 0.05] & 0.01 [-0.08, 0.09] & 0.22 [0.15, 0.3]\\
\hspace{1em}FG-Approx & -0.03 [-0.12, 0.05] & 0.01 [-0.08, 0.09] & 0.22 [0.15, 0.3]\\
\addlinespace[0.3em]
\multicolumn{4}{l}{\textbf{Patient sex: male}}\\
\hspace{1em}CCA & -0.24 [-0.56, 0.09] & -0.18 [-0.51, 0.15] & 0.39 [0.11, 0.68]\\
\hspace{1em}CS-SMC & -0.1 [-0.24, 0.05] & -0.06 [-0.21, 0.09] & 0.18 [0.05, 0.31]\\
\hspace{1em}CS-Approx & -0.1 [-0.24, 0.05] & -0.06 [-0.21, 0.09] & 0.18 [0.05, 0.31]\\
\hspace{1em}FG-SMC & -0.09 [-0.24, 0.05] & -0.06 [-0.2, 0.09] & 0.18 [0.05, 0.31]\\
\hspace{1em}FG-Approx & -0.1 [-0.24, 0.05] & -0.06 [-0.21, 0.08] & 0.18 [0.05, 0.31]\\
\addlinespace[0.3em]
\multicolumn{4}{l}{\textbf{PB Blasts (per $5$\%)}}\\
\hspace{1em}CCA & 0.16 [-0.04, 0.36] & 0.17 [-0.02, 0.37] & 0 [-0.18, 0.18]\\
\hspace{1em}CS-SMC & 0.18 [0.05, 0.31] & 0.18 [0.05, 0.31] & 0.01 [-0.12, 0.13]\\
\hspace{1em}CS-Approx & 0.19 [0.07, 0.31] & 0.19 [0.07, 0.32] & 0.01 [-0.12, 0.13]\\
\hspace{1em}FG-SMC & 0.17 [0.04, 0.3] & 0.17 [0.05, 0.3] & -0.01 [-0.12, 0.1]\\
\hspace{1em}FG-Approx & 0.18 [0.05, 0.32] & 0.18 [0.05, 0.31] & -0.02 [-0.12, 0.09]\\
\addlinespace[0.3em]
\multicolumn{4}{l}{\textbf{Ruxolitinib given: yes}}\\
\hspace{1em}CCA & 0.08 [-0.26, 0.43] & 0.08 [-0.26, 0.43] & -0.05 [-0.33, 0.23]\\
\hspace{1em}CS-SMC & -0.02 [-0.2, 0.17] & -0.03 [-0.22, 0.16] & -0.06 [-0.21, 0.1]\\
\hspace{1em}CS-Approx & 0.01 [-0.19, 0.2] & -0.01 [-0.2, 0.18] & -0.05 [-0.21, 0.11]\\
\hspace{1em}FG-SMC & -0.02 [-0.21, 0.17] & -0.03 [-0.22, 0.16] & -0.04 [-0.19, 0.11]\\
\hspace{1em}FG-Approx & 0 [-0.19, 0.18] & -0.01 [-0.2, 0.17] & -0.04 [-0.19, 0.11]\\
\addlinespace[0.3em]
\multicolumn{4}{l}{\textbf{T-cell depletion: yes}}\\
\hspace{1em}CCA & 0.2 [-0.21, 0.62] & 0.16 [-0.25, 0.58] & -0.23 [-0.54, 0.08]\\
\hspace{1em}CS-SMC & 0.3 [0.13, 0.48] & 0.26 [0.09, 0.44] & -0.18 [-0.32, -0.04]\\
\hspace{1em}CS-Approx & 0.3 [0.12, 0.48] & 0.26 [0.08, 0.43] & -0.19 [-0.33, -0.05]\\
\hspace{1em}FG-SMC & 0.31 [0.13, 0.48] & 0.26 [0.09, 0.44] & -0.18 [-0.31, -0.04]\\
\hspace{1em}FG-Approx & 0.31 [0.13, 0.48] & 0.26 [0.09, 0.44] & -0.18 [-0.32, -0.04]\\
\addlinespace[0.3em]
\multicolumn{4}{l}{\textbf{WBC count (log)}}\\
\hspace{1em}CCA & 0.17 [0.02, 0.33] & 0.17 [0.01, 0.33] & 0.02 [-0.12, 0.15]\\
\hspace{1em}CS-SMC & 0.17 [0.09, 0.26] & 0.18 [0.09, 0.27] & 0 [-0.07, 0.07]\\
\hspace{1em}CS-Approx & 0.17 [0.08, 0.26] & 0.17 [0.09, 0.26] & 0 [-0.08, 0.07]\\
\hspace{1em}FG-SMC & 0.17 [0.09, 0.26] & 0.18 [0.09, 0.26] & -0.01 [-0.07, 0.05]\\
\hspace{1em}FG-Approx & 0.17 [0.1, 0.25] & 0.18 [0.1, 0.26] & -0.01 [-0.08, 0.05]\\
\addlinespace[0.3em]
\multicolumn{4}{l}{\textbf{Weight loss: yes}}\\
\hspace{1em}CCA & 0 [-0.37, 0.38] & 0.05 [-0.33, 0.43] & 0.17 [-0.13, 0.48]\\
\hspace{1em}CS-SMC & 0.23 [-0.03, 0.49] & 0.27 [0.01, 0.53] & 0.16 [-0.05, 0.36]\\
\hspace{1em}CS-Approx & 0.24 [0, 0.47] & 0.28 [0.04, 0.51] & 0.16 [-0.05, 0.36]\\
\hspace{1em}FG-SMC & 0.23 [-0.01, 0.47] & 0.24 [0.01, 0.48] & 0.06 [-0.12, 0.24]\\
\hspace{1em}FG-Approx & 0.24 [0, 0.48] & 0.26 [0.02, 0.49] & 0.06 [-0.14, 0.26]\\
\addlinespace[0.3em]
\multicolumn{4}{l}{\textbf{Year of alloHCT (decades)}}\\
\hspace{1em}CCA & -0.36 [-0.99, 0.26] & -0.41 [-1.04, 0.23] & -0.15 [-0.67, 0.37]\\
\hspace{1em}CS-SMC & -0.08 [-0.34, 0.18] & -0.11 [-0.37, 0.15] & -0.24 [-0.46, -0.02]\\
\hspace{1em}CS-Approx & -0.09 [-0.35, 0.17] & -0.12 [-0.38, 0.14] & -0.24 [-0.46, -0.02]\\
\hspace{1em}FG-SMC & -0.08 [-0.34, 0.17] & -0.12 [-0.37, 0.14] & -0.24 [-0.46, -0.03]\\
\hspace{1em}FG-Approx & -0.08 [-0.34, 0.17] & -0.11 [-0.37, 0.14] & -0.24 [-0.46, -0.03]\\*
\end{longtable}
\endgroup{}